\documentclass[
    aip,
    rsi,
    preprint,
    article-title,      % Explicitly forces article titles in references
    longbibliography,   % Required by REVTeX to output full citation metadata (titles)
    amsmath,amssymb,
    floats,
    floatfix
]{revtex4-1}

\usepackage{xcolor}
\usepackage{float}
\usepackage{graphicx}% Include figure files
\usepackage{dcolumn}% Align table columns on decimal point
\usepackage{bm}% bold math
\usepackage[utf8]{inputenc}
\usepackage[T1]{fontenc}
\usepackage{mathptmx}
\usepackage{etoolbox}
\usepackage{hyperref}
\hypersetup{
  colorlinks=true,
  linkcolor=red,
  citecolor=blue,
  urlcolor=blue
}

\makeatletter
\def\@email#1#2{%
 \endgroup
 \patchcmd{\titleblock@produce}
  {\frontmatter@RRAPformat}
  {\frontmatter@RRAPformat{\produce@RRAP{*#1\href{mailto:#2}{#2}}}\frontmatter@RRAPformat}
  {}{}
}%
\makeatother

\newcommand{\corrauthor}{\textsuperscript{*}}

\begin{document}
\raggedbottom

\title{Micrometer-thin liquid sheets from single-jet impact on a sharp edge for multi-kHz laser–plasma experiments}

\author{Mukesh Kumar}
\affiliation{%
Tata Institute of Fundamental Research Hyderabad, 36/P Gopanpally, Hyderabad 500046, Telangana, India
}

\author{S.~Sree Harsha}
\affiliation{%
Tata Institute of Fundamental Research Hyderabad, 36/P Gopanpally, Hyderabad 500046, Telangana, India
}

\author{Prashant Kumar Singh\corrauthor}
\email{pks@tifrh.res.in}
\affiliation{%
Tata Institute of Fundamental Research Hyderabad, 36/P Gopanpally, Hyderabad 500046, Telangana, India
}

\date{\today}

\begin{abstract}
We present a simple and robust method for generating freely suspended liquid sheets by directing a single laminar liquid jet onto the sharp edge of an inclined plate. By optimizing liquid composition, impact geometry, and flow conditions, we identify a 60:40 ethylene-glycol--deionized-water volumetric mixture for stable sheet formation. The resulting sheets span $8\,\mathrm{mm}$ in height and $3\,\mathrm{mm}$ in width, with thicknesses from $1.1\,\mu\mathrm{m}$ to $4.8\,\mu\mathrm{m}$ measured by spatial and spectral interferometry. Operated in vacuum as continuously renewed plasma mirrors, these targets demonstrate stable performance from $0.5\,\mathrm{kHz}$ to $10\,\mathrm{kHz}$ using $175\text{-}\mathrm{fs}$ Yb:KGW laser pulses, reaching a maximum $S$-polarized reflectivity of $66.5\%$ at an intensity of $4\times10^{14}\,\mathrm{W\,cm^{-2}}$. At higher intensities using $25\text{-}\mathrm{fs}$ Ti:sapphire pulses at $1\,\mathrm{kHz}$, maximum $S$-polarized reflectivity reaches $77\%$ at $2\times10^{16}\,\mathrm{W\,cm^{-2}}$. The reflected beam exhibits a high temporal stability with a angular pointing fluctuations of $18.19 \pm 0.53\,\mu\mathrm{rad}$. Our results demonstrate a tunable, continuously renewed liquid-sheet target for multi-kHz laser--plasma experiments.
\end{abstract}

\maketitle

% Your main text starts here. Every line will be numbered because of \linenumbers.
% Your main text starts here.

\section{Introduction}
The study of laser--plasma interactions involves subjecting targets to extreme conditions generated by high-intensity laser pulses, under which the irradiated region is permanently altered or destroyed~\cite{kruer2019physics,gibbon1996short}. High-repetition-rate operation therefore requires the target to be replenished or replaced before each subsequent shot, demanding a well-controlled and reproducible initial state for every shot~\cite{george2019high,ngirmang2020evidence}. Bulk solid targets, although widely used, are poorly suited to high-repetition-rate experiments, since each shot destroys the target and generates debris from its surface~\cite{ehret2024stability}. Gas jets are renewable but lack sharp density gradients and are often restricted in achievable density~\cite{10.1063/1.5093613,dover2016optical}. Liquid-sheet targets bridge this gap. Typically produced by colliding microjets or using microfluidic nozzles, liquid sheets provide a flat, optically smooth interaction surface that is continuously renewed between laser shots~\cite{dombrowski1954photographic,ha2018device}, making them promising targets for high-repetition-rate laser--plasma experiments~\cite{dombrowski1954photographic,ha2018device,10.1063/5.0328950,10.1063/5.0279613}. In addition, liquid sheets combine a solid-like mass density, a thickness tunable from tens of nanometers to tens of micrometers~\cite{koralek2018generation,ekimova2015liquid,hoffman2022liquid,menzi2020generation}, and optically smooth interfaces~\cite{peng2024comprehensive,gallo2024development,crissman2022sub}. This combination is difficult to achieve with conventional solids~\cite{prencipe2017targets,Borghesi01042006,borot2014high}, thin films~\cite{schumacher2017liquid}, or tape targets~\cite{condamine2021high,prencipe2017targets,xu2023versatile}, and makes liquid sheets particularly well suited to next-generation, multi-kHz laser systems. Liquid sheets have been used as targets for plasma mirrors~\cite{underwood2020characterization}, high-harmonic and extreme-ultraviolet generation~\cite{kim2023high,luu2018extreme,10.1063/5.0316700}, electron and ion acceleration~\cite{morrison2018mev,ggsf-hkm1,treffert2022high,streeter2025stable,puyuelo2022implementation,PhysRevResearch.7.023190}, and laser-driven x-ray and neutron sources~\cite{knight2024detailed,hah2016high,kar2008plasma}. Although liquid targets also generate vapour and ejected liquid, established approaches to vapour and liquid management include cryogenic trapping~\cite{faubel1988molecular,barnard2022delivery,riley2019design}, differential pumping~\cite{george2019high}, and closed-loop recirculation~\cite{ekimova2015liquid,galinis2017micrometer}. Realizing the full potential of liquid-sheet targets at high repetition rates, however, requires generation methods that are simple, robust, and compatible with vacuum operation.

Liquid-sheet formation generally involves the conversion of the directed momentum of one or more liquid streams into lateral \cite{bremond2006atomization}. When two liquid jets collide, their axial momenta are redirected into the plane normal to the collision axis, producing a radially expanding sheet\cite{sanjay2017formation}. The liquid accumulates at the perimeter, where surface tension creates a thicker rim that confines the sheet\cite{bush2004collision}. As the sheet expands, its thickness decreases because approximately the same volumetric flow is distributed over an increasing area\cite{hasson1964thickness}. The competition between liquid inertia and capillary forces, commonly characterized by the Weber number, determines the sheet extent, thickness, rim dynamics, and eventual breakup\cite{li2006characteristics}.

The same general balance between inertia and surface tension appears in other sheet-generation geometries. A liquid jet impacting a solid surface spreads radially after the stagnation region\cite{yan1997heat,Kaewchoothong31052026,mishra2020experimental}, whereas a slit or shaped nozzle directly produces a planar sheet by imposing a two-dimensional exit geometry\cite{ha2018device}. Gas-assisted and microfluidic devices further modify the sheet through gas momentum, channel confinement, or controlled interaction of multiple liquid streams. Based on these formation mechanisms, liquid sheets have been generated using colliding liquid jets~\cite{dombrowski1954photographic,ha2018device,taylor1961formation,bush2004collision}, slit and microfluidic nozzles~\cite{PhysRevAccelBeams.24.121301,article,ohba2014effect}, vibration-assisted devices~\cite{cao2023vibration}, and splash-plate geometries~\cite{ahmed2008characteristics}. These approaches provide different levels of control over sheet thickness and lateral dimension\cite{10.1039/d1lc00757b}, but some require specialized nozzle fabrication, multiple interacting flows, or precise positioning of micrometer-scale flow \cite{koralek2018generation}. Such requirements can complicate integration with vacuum chambers and high-repetition-rate laser systems, and they motivate the development of simpler, more robust sheet-generation methods that retain tunability and stability.

Here, we use a momentum-redirecting geometry in which a single liquid jet impacts the sharp edge of an inclined plate. Whereas normal impact on a planar surface produces a radially spreading film after the stagnation region~\cite{nawani2021review,kim2020experimental,10.1115/1.2911284,watson1964radial}, oblique impact generates an asymmetric spreading flow~\cite{taylor1966oblique,jalil2006oblique,kate2007hydraulic,10.1063/5.0047568}. Interruption of this flow at a sharp edge produces a freely suspended liquid sheet downstream. Unlike conventional liquid sheets generated by colliding two liquid jets~\cite{yang2014liquid}, our approach uses a single jet impinging on the sharp edge of an inclined plate. This simplifies alignment because the jet needs to be positioned relative to the plate edge, rather than aligned with a second jet to establish a collision point. We systematically investigate the effects of liquid composition, jet--plate angle, impact distance from the edge, and jet velocity on sheet formation and morphology~\cite{choo2002velocity,zhao2016characteristics,weng2025experimental}. Among the tested liquids, a 60:40 ethylene-glycol--deionized-water mixture by volume produces the most stable sheet. Characterization of the blade surface and edge reveals micrometer-scale machining grooves that may contribute to sheet wrinkling. Using complementary spectral and spatial interferometry, we measure sheet thicknesses ranging from $1.1$ to $4.8\,\mu\mathrm{m}$ and compare the measured sheet shape and thickness profile with established analytical models~\cite{hasson1964thickness,ibrahim1991impinging,li2006characteristics}. These measurements establish the geometric and thickness characteristics of the sheet under the selected operating conditions.

We then demonstrate the use of this continuously renewed sheet as a plasma mirror~\cite{levy2007double,vincenti2014optical}. Using $175$-fs Yb:KGW pulses, we investigate reflection and transmission as functions of incident intensity and repetition rate, demonstrating operation up to $10$~kHz. The maximum measured $S$-polarized reflectivity is $66.5\%$ at an intensity of approximately $4\times10^{14}\,\mathrm{W\,cm^{-2}}$. Complementary measurements with $25$-fs Ti:sapphire pulses extend the characterization to higher intensities, yielding a maximum $S$-polarized reflectivity of $77\%$ at approximately $2\times10^{16}\,\mathrm{W\,cm^{-2}}$. Near- and far-field imaging, spectral measurements, and temporal tracking of the reflected beam characterize its optical response and stability, including an angular pointing fluctuation of $18.19\pm0.53\,\mu\mathrm{rad}$ during operation at $10$~kHz. These results establish the potential of this geometry for high-repetition-rate laser--plasma experiments and associated secondary-radiation applications~\cite{clark2021high}.

%%%%%%%%%%%%%%%%%%%%%%%%%%%%%%%%%%%%%%%%%%%%%%%%%%%%%%%%%%%%%%%%%%%%%%%%%%%%%%%%%%%%%%%%%%%%%%%%%%%%%%%%%%%%%%%%%%%%%%%%%%%%%%%%%%%%%%%%%%%%%%%%%%%%%%%%%%%%%%%%%%%%%%%%%%%%%%%%%%%%%%%%%%%%%%%%%%%%%%%%%%%%%%%%%%%%%%%%%%%%%%%%%%%%%%%%%%%%%%%%%%%%%%%%%%%%%%%%%%%%%%%%%%%%%%%%%%%%%%%%%%%%%%%%%%%%%%%%%%%%%%%%%%%%%%%%%%%%%%%%%%%%%%%%%%%%%%%%%%%%%%%%%%%%%%%%%%%%%%%%
\section{Experimental setup and generation of liquid sheet}

The formation of a liquid sheet by jet impact depends on how the incident momentum is redistributed by the impact geometry. For normal impact on a broad planar surface, the incoming liquid spreads radially from the stagnation region, producing an approximately axisymmetric film~\cite{nawani2021review,kim2020experimental,10.1115/1.2911284,watson1964radial}. Oblique impact breaks this symmetry by introducing a component of momentum parallel to the surface, resulting in an asymmetric, approximately elliptical spreading flow~\cite{taylor1966oblique,jalil2006oblique,kate2007hydraulic,10.1063/5.0047568}. When this spreading flow encounters a sharp boundary, a portion of the liquid leaves the supporting surface and forms a freely suspended sheet.

\begin{figure}[!htbp]
    \centering
    \includegraphics[width=1\linewidth]{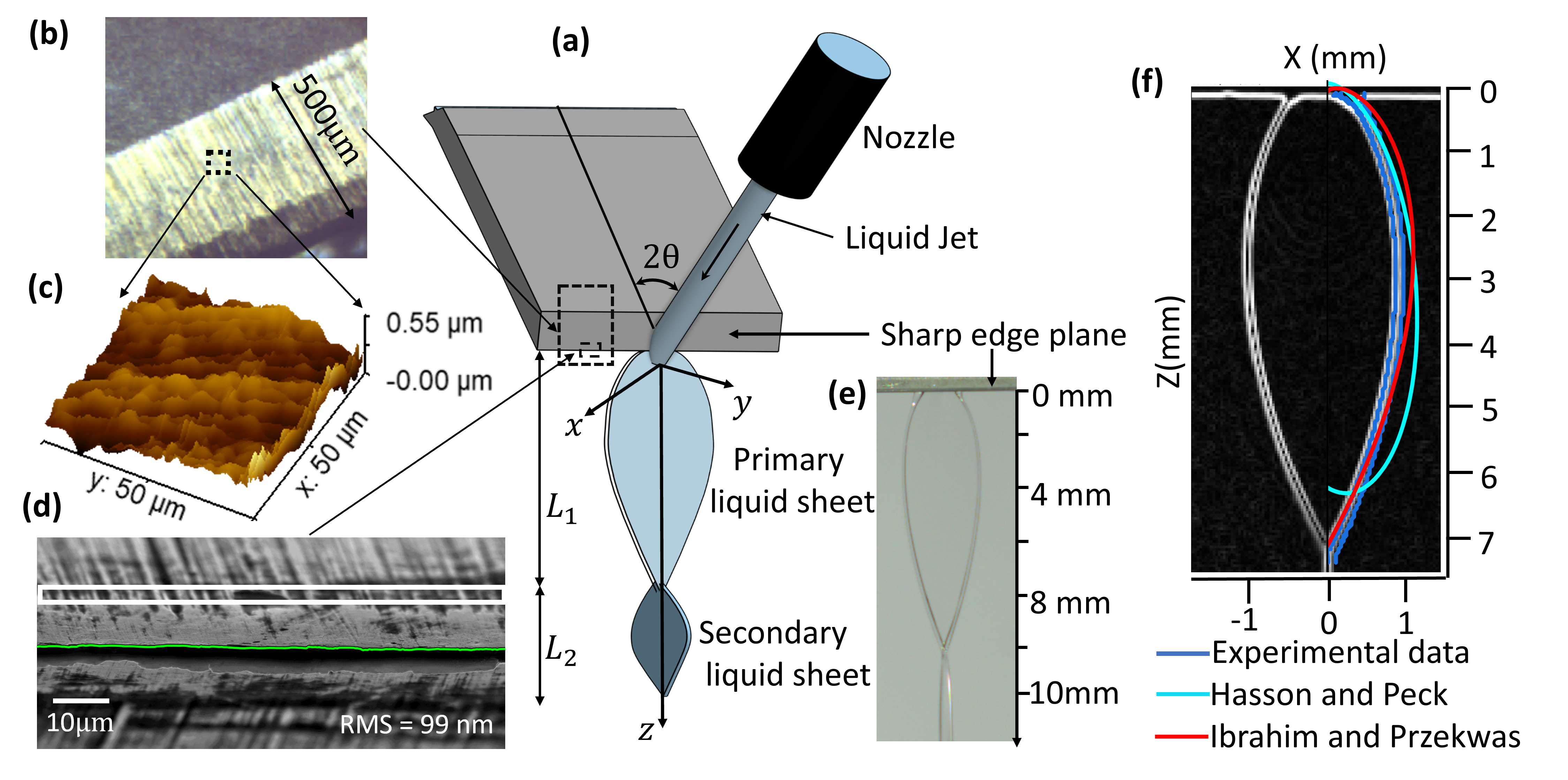}
    \caption{\textbf{Formation and structure of liquid sheets generated by inclined jet impact on a sharp razor-blade edge.}
    \textbf{(a)} Concept of liquid-sheet formation by impact of a single liquid jet on the inclined sharp edge of a stainless-steel razor blade. The inclination angle $2\theta$ is defined between the incoming jet axis and the sharp edge. $L_1$ and $L_2$ denote the lengths of the primary and secondary sheets, which extend approximately orthogonally.
    \textbf{(b)} Optical-microscope image of the razor-blade edge and the jet-impact region selected for surface characterization.
    \textbf{(c)} Atomic force microscopy (AFM) topography of a $50\,\mu\mathrm{m}\times50\,\mu\mathrm{m}$ region at the jet-impact location, showing preferentially oriented machining grooves with an rms surface roughness of $41$~nm.
    \textbf{(d)} Scanning electron microscopy (SEM) images of the sharp blade edge. Green curves trace the extracted edge contour, with an rms deviation of $100$~nm. The white rectangle marks the jet-impact region used to quantify the machining grooves, giving a groove width of $1.14\,\mu\mathrm{m}$ and an inter-groove spacing of $2.2\,\mu\mathrm{m}$. Panels (b)--(d) characterize the blade-edge geometry and microscopic surface structure encountered by the impacting jet.
    \textbf{(e)} Representative liquid sheet generated using a 60:40 ethylene glycol--deionized-water mixture by volume at a jet velocity of $13$~m\,s$^{-1}$ and $2\theta=64\pm1^\circ$. The primary and secondary sheets are indicated.
    \textbf{(f)} Comparison of the measured sheet dimensions with the theoretical model for $We=271\pm42$, $2\theta=64\pm1^\circ$, and a jet diameter of $100\,\mu\mathrm{m}$.}
    \label{fig:placeholder}
\end{figure}

Our experimental configuration uses a single laminar liquid jet impinging on the sharp edge of a commercial stainless-steel razor blade inclined at an angle to the jet axis [Fig.~1(a)], rather than the collision of two liquid jets commonly used to generate liquid sheets~\cite{yang2014liquid}. The resulting sheet comprises primary and secondary branches, with lengths $L_1$ and $L_2$, respectively. Its geometry and stability depend on the inclination angle $2\theta$ and the distance $d$ between the jet impact point and the sharp edge~\cite{choo2002velocity,zhao2016characteristics,weng2025experimental}. The influence of these parameters on sheet formation is examined in detail in the following section.

The blade surface topography and edge straightness may influence the uniformity of the emerging liquid sheet. We therefore characterized both near the jet-impact region using optical microscopy, atomic force microscopy (AFM), and scanning electron microscopy (SEM) [Figs.~1(b)--1(d)]. The optical micrograph in Fig. 1(b) reveals that the blade surface near the jet-impact region is not perfectly flat and exhibits visible machining marks. These surface imperfections were characterized using atomic force microscopy (AFM), while the edge contour and machining grooves were further examined using scanning electron microscopy (SEM) [Figs. 1(c) and 1(d)]. The AFM topography in Fig.~1(c), acquired over a $50\,\mu\mathrm{m}\times50\,\mu\mathrm{m}$ region, reveals preferentially oriented machining grooves and an RMS surface roughness of $41\,\mathrm{nm}$. Analysis of the SEM images in Fig.~1(d) gives an RMS deviation of approximately $100 \mathrm{nm}$ for the extracted edge contour over a projected length of $80 \mu\mathrm{m}$. This deviation is only $0.1\%$ of the $100 \mu\mathrm{m}$ jet diameter, indicating that the edge contour is well defined at the scale of the incident jet. The SEM image also resolves machining grooves with a characteristic width of $1.14\ \mu\mathrm{m}$ and an inter-groove spacing of $2.2\,\mu\mathrm{m}$. These dimensions correspond to approximately $1.1\%$ and $2.2\%$ of the jet diameter, respectively. Although the edge-profile deviation is much smaller than the jet diameter, the micrometre-scale grooves could perturb the flow as the jet spreads across the plate and is redirected at the sharp edge, potentially producing wrinkles in the otherwise smooth, freely suspended sheet. The influence of such disturbances may be mitigated by using more viscous liquids, such as ethylene glycol--water mixtures. The role of liquid properties in determining sheet morphology and the optimization of mixture composition are discussed in the following section.

Using a mixture of $60\%$ ethylene glycol and $40\%$ deionized (DI) water by volume, we obtained a freely suspended liquid sheet from a $100\,\mu\mathrm{m}$-diameter jet, as shown in Fig.~1(e). The representative sheet exhibits distinct primary and secondary branches generated by the single-jet impact configuration. We further quantified the resulting sheet dimensions and compared them with the theoretical model [Fig.~1(f)]. This comparison provides a quantitative assessment of the model's ability to describe the observed sheet geometry under the experimental conditions.

%%%%%%%%%%%%%%%%%%%%%%%%%%%%%%%%%%%%%%%%%%%%%%%%%%%%%%%%%%%%%%%%%%%%%%%%%%%%%%%%%%%%%%%%%%
%%%%%%%%%%%%%%%%%%%%%%%%%%%%%%%%%%%%%%%%%%%%%%%%%%%%%%%%%%%%%%%%%%%%%%%%%%%%%%%%%%%%%%%%%%
%%%%%%%%%%%%%%%%%%%%%%%%%%%%%%%%%%%%%%%%%%%%%%%%%%%%%%%%%%%%%%%%%%%%%%%%%%%%%%%%%%%%%%%%%%
%%%%%%%%%%%%%%%%%%%%%%%%%%%%%%%%%%%%%%%%%%%%%%%%%%%%%%%%%%%%%%%%%%%%%%%%%%%%%%%%%%%%%%%%%%

\section{Agreement between theory and experiment for the liquid-sheet shape}

We compare the measured liquid-sheet boundary with two analytical models developed for impinging-jet sheets: the Hasson--Peck model~\cite{hasson1964thickness} and the Ibrahim--Przekwas model~\cite{ibrahim1991impinging}. Hasson and Peck assumed an elliptical impingement region and obtained the following expression for the sheet extent~\cite{li2006characteristics}:

\begin{equation}
\frac{r_e}{R\mathrm{We}} =
\frac{\sin^3\theta\sin^2\psi}
{4(1-\cos\phi\cos\theta)^2}.
\label{eq:1}
\end{equation}

Ibrahim and Przekwas instead assumed a circular impingement region and obtained~\cite{li2006characteristics}

\begin{equation}
\frac{r_e}{R\mathrm{We}} =
\frac{\beta e^{\beta(1-\phi/\pi)}\sin^2\psi}
{4(e^\beta-1)}.
\label{eq:2}
\end{equation}

Here, $R$ is the jet radius, $\mathrm{We}$ is the Weber number, $\theta$ and $\phi$ define the impingement geometry, $\psi$ is a function of $\theta$ and $\phi$, and $\beta$ is a parameter that depends on $\theta$ and controls the azimuthal variation of the predicted sheet extent. Both expressions were transformed from polar to Cartesian coordinates for direct comparison with the measured sheet contour.

Figure~1(f) shows the raw image of the sheet rim together with the experimentally extracted contour (blue) and the predictions of the Hasson--Peck (cyan) and Ibrahim--Przekwas (red) models. Equation~(\ref{eq:2}) agrees more closely with the measured contour than equation~(\ref{eq:1}), reproducing the sheet boundary at $z=0$~mm and near $z=7$~mm. The improved agreement is consistent with the circular-impingement assumption and the $\beta$-dependent azimuthal correction in the Ibrahim--Przekwas model, which allows the predicted sheet curvature to vary more strongly with azimuthal angle. Nevertheless, the model does not fully reproduce the sheet tip near $z=7$~mm, indicating that the sharp-edge impact geometry introduces features not captured by models developed for conventional impinging-jet sheets. For the conditions examined here, the Ibrahim--Przekwas model therefore provides the closer description of the measured sheet morphology, particularly in regions of strong curvature.

%%%%%%%%%%%%%%%%%%%%%%%%%%%%%%%%%%%%%%%%%%%%%%%%%%%%%%%%%%%%%%%%%%%%%%%%%%%%%%%%%%%%%%%%%%
%%%%%%%%%%%%%%%%%%%%%%%%%%%%%%%%%%%%%%%%%%%%%%%%%%%%%%%%%%%%%%%%%%%%%%%%%%%%%%%%%%%%%%%%%%
%%%%%%%%%%%%%%%%%%%%%%%%%%%%%%%%%%%%%%%%%%%%%%%%%%%%%%%%%%%%%%%%%%%%%%%%%%%%%%%%%%%%%%%%%%
%%%%%%%%%%%%%%%%%%%%%%%%%%%%%%%%%%%%%%%%%%%%%%%%%%%%%%%%%%%%%%%%%%%%%%%%%%%%%%%%%%%%%%%%%%

\section{Optimization of liquid-sheet generation}
\label{sec:optimization}

\begin{figure}[h!]
    \centering
    \includegraphics[width=1\linewidth]{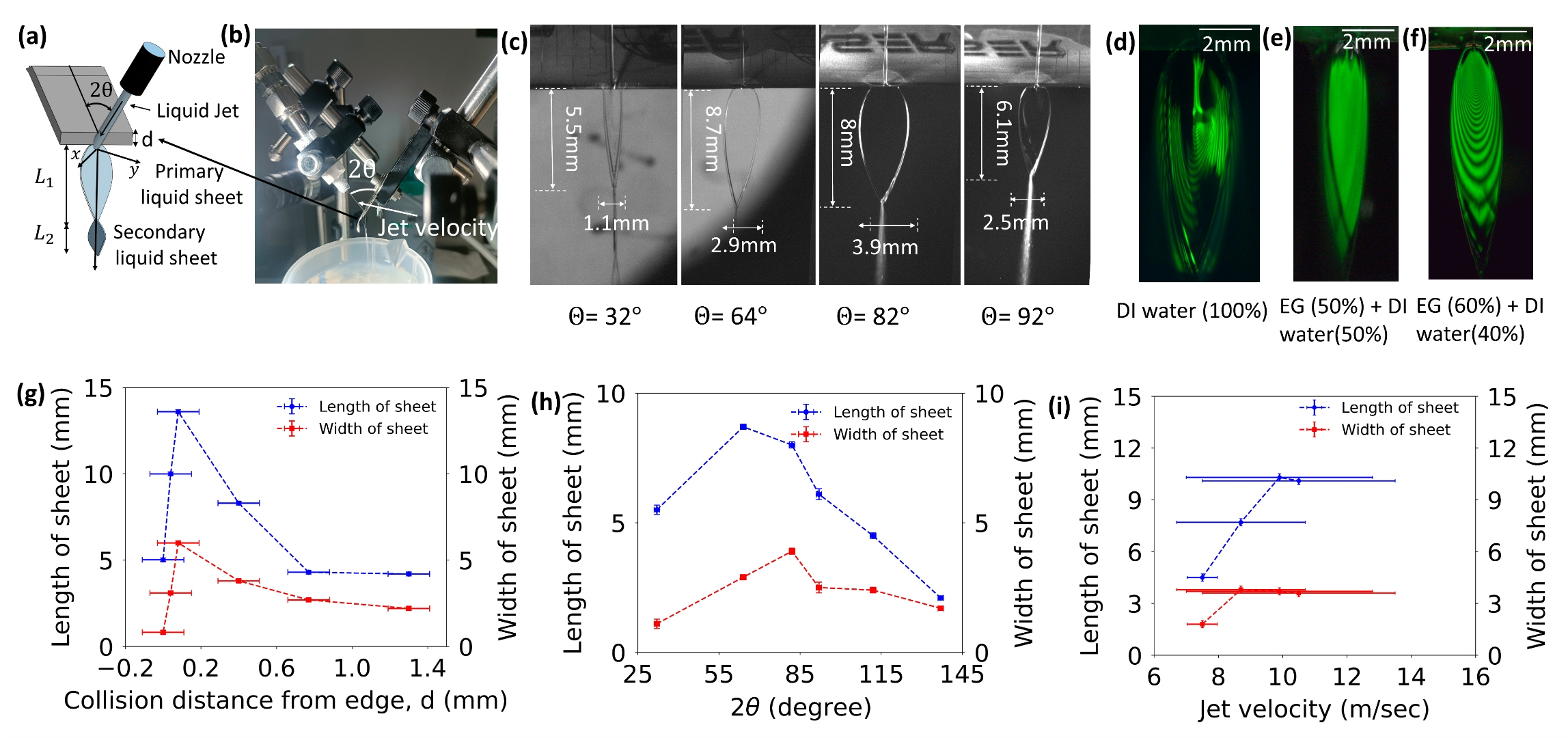}
    \caption{\textbf{Optimization of liquid-sheet generation by liquid properties, impact geometry, and flow conditions.}
    \textbf{(a)} Schematic of the sheet-generation geometry, showing the angle between the jet axis and the inclined plate, the distance between the jet impact point and the sharp edge, the jet velocity, and the liquid composition as the principal control parameters.
    \textbf{(b)} Photograph of the experimental setup.
    \textbf{(c)} Representative liquid-sheet images obtained for selected jet--plate angles.
    \textbf{(d)--(f)} Interferometric images of sheets generated using \textbf{(d)} deionized water, \textbf{(e)} a 50:50 ethylene glycol--deionized-water mixture by volume, and \textbf{(f)} a 60:40 ethylene glycol--deionized-water mixture by volume.
    \textbf{(g)--(i)} Dependence of the liquid-sheet length and width on \textbf{(g)} the jet impact distance from the sharp edge, \textbf{(h)} the angle between the jet axis and the inclined plate, and \textbf{(i)} the jet velocity.}
    \label{fig:optimization}
\end{figure}

The liquid-sheet morphology is governed by the liquid density $\rho$, dynamic viscosity $\mu$, surface tension $\gamma$, jet velocity, jet diameter, and impact geometry. These parameters determine the relative contributions of inertia, viscous dissipation, and capillary forces during jet impact, spreading, rim formation, and sheet breakup. We therefore investigated the effects of liquid composition, the angle between the jet and the inclined plate, the distance between the jet impact point and the sharp edge, and the jet velocity. The principal control parameters and the corresponding sheet measurements are summarized in Fig.~\ref{fig:optimization}.

\subsection{Dependence on the jet--plate angle}

We first investigated the effect of the angle $2\theta$ between the liquid-jet axis and the inclined plate. The jet diameter for this scan was $237\,\mu\mathrm{m}$. The jet impacted the flat portion of the plate, upstream of the sharp edge; therefore, $2\theta$ defines the angle of jet incidence on the plate rather than an angle between the jet and the edge. Owing to the mechanical constraints imposed by the optomechanical assembly, the accessible angular range was $2\theta=32^\circ$--$137^\circ$.

At each angle, the liquid first spread over the flat plate and subsequently reached the sharp edge. At $2\theta=32^\circ$, the resulting sheet was narrow, elongated, and leaf-shaped. Although a sheet was formed, its orientation was not suitable for the intended laser--plasma interaction geometry. As the angle increased, the sheet became wider and developed a more pronounced leaf-like shape because the spreading flow encountered the sharp edge with a different momentum distribution. The corresponding changes in sheet length and width are shown in Fig.~\ref{fig:optimization}(h).

At larger angles, the jet approached the plate more directly, and the flow reaching the edge produced a more irregular and dynamically disturbed sheet. Within the accessible range, the most useful operating window was $2\theta=50^\circ$--$85^\circ$. In this range, the sheet was sufficiently stable and approximately perpendicular to the desired interaction plane for use as a liquid target. The angle scan therefore identified $2\theta=50^\circ$--$85^\circ$ as the practical operating range for the subsequent experiments.

\subsection{Dependence on the jet impact distance from the sharp edge}

The impact distance $d$ determines how the liquid spreading and rim formation interact with the sharp edge. When the jet impacts too close to the edge, the available plate surface is insufficient for the liquid to spread and redistribute before reaching the edge, limiting the formation of a freely suspended sheet. As $d$ increases, the spreading flow develops over a larger surface area, allowing a greater fraction of the liquid rim to extend beyond the edge and increasing the sheet dimensions. However, when the impact point is too far from the edge, much of the spreading and rim formation occurs on the plate itself. The rim may approach or complete a closed loop before reaching the edge, leaving less liquid to form the suspended sheet. We therefore expect the sheet dimensions to reach a maximum at an intermediate impact distance.

To test this expectation, we varied $d$ by translating the razor blade while keeping the jet fixed, with $d=0$ defined at the bottom sharp edge [Fig.~2(a)]. The measurements in Fig.~2(g) show that the sheet length increases from approximately $5$~mm at $d\approx-0.05$~mm to $10$~mm at $d\approx0.05$~mm, reaching a maximum of approximately $13.5$~mm near $d=0.1$~mm. The sheet width follows a similar trend, increasing from less than $1$~mm to approximately $6$~mm at the same optimal distance. At larger impact distances, both dimensions decrease: the sheet length falls to approximately $8.3$~mm at $d=0.4$~mm and $4.3$~mm at $d=0.75$~mm, while the width decreases to approximately $3.8$ and $2.7$~mm, respectively. At the largest measured distances, the length remains near $4.2$~mm, whereas the width decreases further to approximately $2.2$~mm.

Thus, the largest sheet is obtained near $d=0.1$~mm. Considering both sheet dimensions and the observed sheet morphology, we identify the intermediate impact-distance region around $d=0.05$--$0.4$~mm as the candidate operating range for generating an extended, freely suspended sheet. The precise usable range depends on the required sheet dimensions and stability.

\subsection{Dependence on jet velocity}

The jet velocity influences liquid-sheet formation by controlling the relative importance of inertial and capillary forces during impact and spreading. This balance is characterized by the Weber number, $We = \frac{\rho U_j^2 D_j}{\gamma}$, where $\rho$ is the liquid density, $U_j$ is the jet velocity, $D_j$ is the jet diameter, and $\gamma$ is the surface tension. Greater inertia generally promotes more extensive spreading after impact. However, the blade surface is not perfectly smooth, and its machining grooves can introduce small perturbations into the spreading flow. At higher velocities, such disturbances may become more pronounced and can potentially grow into larger distortions of the sheet. Consequently, maximizing the jet velocity alone does not necessarily provide the most suitable target for laser--plasma experiments, where a sufficiently extended and uniform interaction region must be maintained during high-repetition-rate operation.

A $237\,\mu\mathrm{m}$-diameter jet was driven by nitrogen gas, and its velocity was varied using a valve in the liquid-delivery line. The velocity was estimated from the collected liquid volume over a measured time interval, $U_j=4V/(\pi D_j^2\Delta t)$, and independently verified using a femtosecond pump--probe technique. As shown in Fig.~2(i), increasing $U_j$ from approximately $7.5$ to $9.8$~m\,s$^{-1}$ increases the sheet length from $4.5$ to $10$~mm and its width from $1.8$ to $3.8$~mm. At $10$--$10.5$~m\,s$^{-1}$, these dimensions remain nearly unchanged. We therefore selected $U_j\approx10$~m\,s$^{-1}$ for subsequent laser--plasma experiments, providing a $10$-mm-long, $3.8$-mm-wide sheet while maintaining the required uniformity and stability.

\subsection{Dependence on liquid composition}

Finally, we varied the liquid composition to identify a stable sheet suitable for use as a laser target. The tested liquids---methanol, deionized water, and ethylene glycol---span substantially different viscosities and surface tensions, thereby affecting jet formation, sheet spreading, and stability. At approximately $20^\circ$C, methanol has a dynamic viscosity of $0.54$~mPa\,s and a surface tension of $22.1$~mN\,m$^{-1}$, while deionized water has a viscosity of $1.00$~mPa\,s and a surface tension of $72.8$~mN\,m$^{-1}$. Ethylene glycol has a substantially higher viscosity of $16.1$~mPa\,s and a surface tension of $48.0$~mN\,m$^{-1}$ \cite{souckova2008measurement,xiang2006new,vargaftik1983international,quijada2013viscosity}.

Methanol spread excessively after impact and produced a less stable sheet that sometimes broke before forming a closed rim. Deionized water produced a more confined sheet, but pronounced wrinkles, irregularities, and spraying remained near the sharp edge [Fig.~\ref{fig:optimization}(d)]. The blade's machining grooves, with a width of approximately $1.1\,\mu\mathrm{m}$ and an inter-groove spacing of $2.2\,\mu\mathrm{m}$ [Fig.~1(d)], may contribute to these wrinkles. Their influence could be reduced by using a smoother blade surface or a more viscous liquid. Consistent with the latter approach, a 50:50 ethylene-glycol--water mixture improved sheet uniformity, although some wrinkles persisted [Fig.~\ref{fig:optimization}(e)]. Increasing the ethylene glycol fraction to 60\% further suppressed wrinkling and produced the most stable sheet among the tested compositions [Fig.~\ref{fig:optimization}(f)]. Pure ethylene glycol, however, was difficult to drive through the $100\,\mu\mathrm{m}$ nozzle at low backing pressure, limiting its suitability for generating a stable, free-flowing jet under our operating conditions.

The 60:40 ethylene-glycol--water mixture by volume was therefore selected for subsequent interferometric thickness measurements and plasma-mirror experiments. Although its interference fringes indicate that the sheet was not globally flat, it provided a sufficiently flat micrometre-scale interaction region for the laser experiments. The angle scan identified $2\theta=50^\circ$--$85^\circ$ as the practical operating range, while the impact-distance and velocity scans established the corresponding geometric and flow conditions.

%%%%%%%%%%%%%%%%%%%%%%%%%%%%%%%%%%%%%%%%%%%%%%%%%%%%%%%%%%%%%%%%%%%%%%%%%%%%%%%%%%%%%%%%%%
%%%%%%%%%%%%%%%%%%%%%%%%%%%%%%%%%%%%%%%%%%%%%%%%%%%%%%%%%%%%%%%%%%%%%%%%%%%%%%%%%%%%%%%%%%
%%%%%%%%%%%%%%%%%%%%%%%%%%%%%%%%%%%%%%%%%%%%%%%%%%%%%%%%%%%%%%%%%%%%%%%%%%%%%%%%%%%%%%%%%%
%%%%%%%%%%%%%%%%%%%%%%%%%%%%%%%%%%%%%%%%%%%%%%%%%%%%%%%%%%%%%%%%%%%%%%%%%%%%%%%%%%%%%%%%%%

\section{Measurement of liquid-sheet thickness by spectral and spatial interferometry}
\label{sec:thickness_measurement}

Following the optimization of the liquid-sheet generation parameters, we measured the thickness of the sheet under the selected operating conditions. The liquid sheet was generated in air using a 60:40 ethylene glycol--deionized-water mixture by volume at a driving pressure of $8$~bar and a jet--edge angle of $2\theta=64\pm1^\circ$. The sheet thickness was measured using two complementary optical techniques: spectral interferometry and spatial interferometry\cite{choo2001parametric}.

\begin{figure}[h!]
    \centering
    \includegraphics[width=1\linewidth]{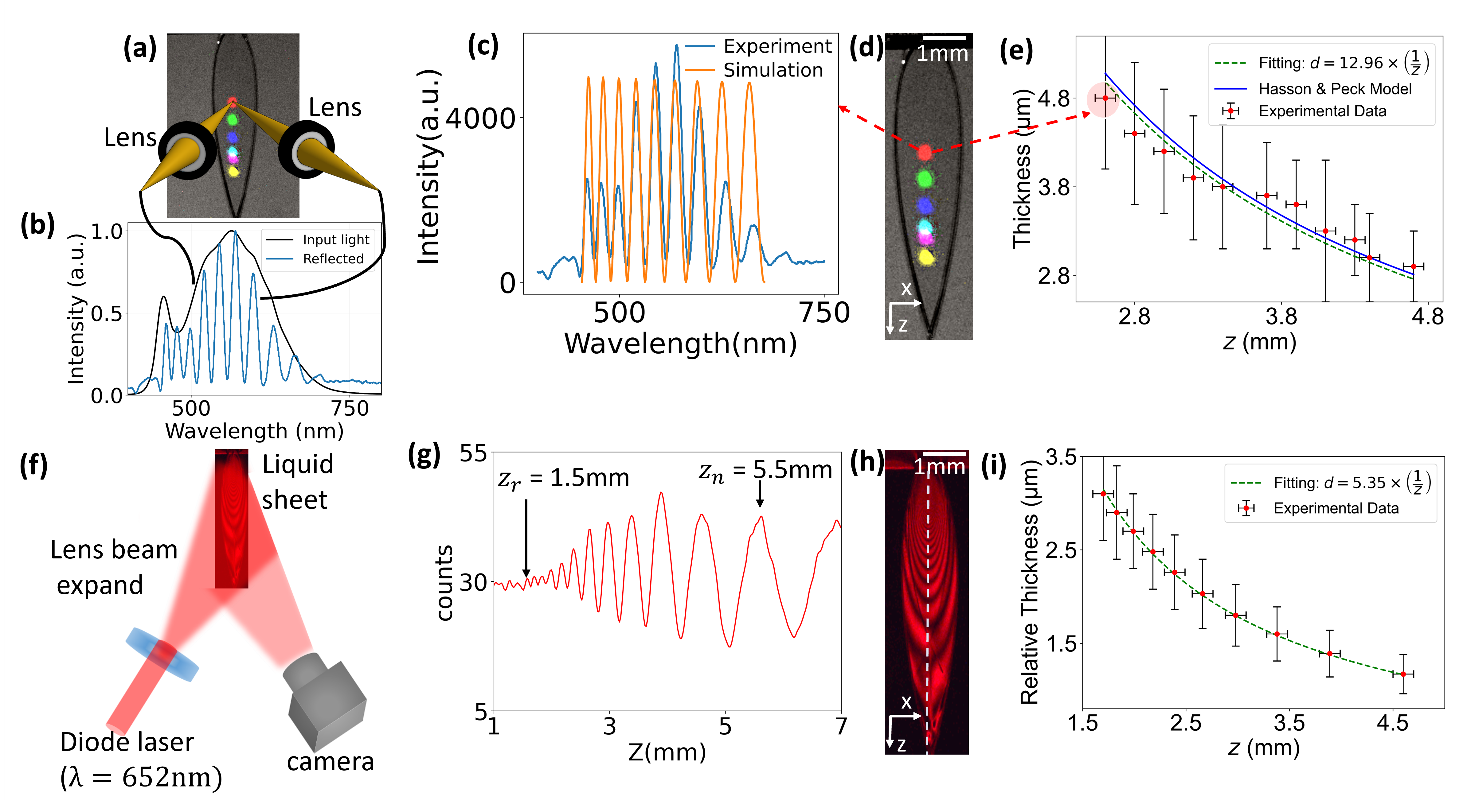}
    \caption{\textbf{Spectral and spatial interferometric measurements of liquid-sheet thickness.}
    \textbf{(a)} Experimental arrangement for measuring the sheet thickness using spectral interferometry. The colored markers indicate the measurement locations on the liquid sheet.
    \textbf{(b)} Incident LED spectrum (black) and spectrum reflected from the liquid sheet (blue).
    \textbf{(c)} Measured reflected spectrum (blue) and the corresponding spectrum calculated using a thin-film interference model (orange).
    \textbf{(d)} Image of the liquid sheet showing the measurement locations.
    \textbf{(e)} Liquid-sheet thickness profile along the sheet-flow direction ($z$). The red markers represent the experimental measurements, the solid curve is a fit to the data, and the blue curve denotes the analytical thickness profile obtained from the Hasson--Peck model.
    \textbf{(f)} Experimental arrangement for measuring the sheet thickness using spatial interferometry.
    \textbf{(g)} Intensity line scan along the sheet-flow direction ($z$), covering the region from $z=1.5$ to $5.5$~mm on the liquid sheet.
    \textbf{(h)} Spatial interference fringes produced by illuminating the liquid sheet with monochromatic light of wavelength $\lambda=652$~nm.
    \textbf{(i)} Relative thickness profile reconstructed from the spatial interference fringes.} 
    \label{fig:thickness}
\end{figure}

Spectral interferometry is an established method for measuring the thickness of transparent liquid films
\cite{watanabe1989new,koralek2018generation,kondoh2014liquid,galinis2017micrometer}. The experimental arrangement is shown in Fig.~\ref{fig:thickness}(a). Light from a broadband LED source with a spectral range of approximately $400$--$750$~nm was focused onto the liquid sheet using a lens. The light reflected from the front and rear interfaces of the sheet was collected and refocused by a second lens. An optical fiber positioned at the focal region coupled the reflected light into a spectrometer (AvaSpec-3648-USB2) for spectral analysis.

The incident LED spectrum and the spectrum reflected from the liquid sheet are shown in Fig.~\ref{fig:thickness}(b). The reflected signal contains contributions from both interfaces of the sheet, and their interference produces a thickness-dependent modulation in the spectrum. Figure~\ref{fig:thickness}(c) shows the measured reflected spectrum together with the spectrum calculated using a thin-film interference model. The agreement between the two spectra allows the local sheet thickness to be extracted from the measured interference pattern.

To determine the spatial variation of the thickness, spectral measurements were performed at multiple positions along the sheet-flow direction, denoted by $z$. The measurement positions are shown in Fig.~\ref{fig:thickness}(d). At each position, the sheet thickness was calculated using
\begin{equation}
d =
\frac{\Delta j}
{4\left(\nu_{j+\Delta j}-\nu_j\right)
\sqrt{n_2^2-\sin^2\phi}},
\label{eq:spectral_thickness}
\end{equation}
where $\Delta j$ is the fringe-order difference, $\nu_j$ and $\nu_{j+\Delta j}$ are the corresponding spectral wavenumbers, $n_2$ is the refractive index of the liquid, and $\phi$ is the angle of incidence
\cite{stenzel2005physics}. The measured thickness values are shown by the red markers in Fig.~\ref{fig:thickness}(e).

The experimental profile was compared with the analytical thickness distribution derived by Hasson and Peck
\cite{hasson1964thickness},
\begin{equation}
d =
\frac{R^2\sin^3\theta}
{r\left(1-\cos\theta\cos\phi\right)^2},
\label{eq:hasson_peck}
\end{equation}
where $R$ is the jet radius, $r$ is the distance from the effective origin of the sheet, and $\theta$ and $\phi$ are the geometric angles defined in Fig.~\ref{fig:thickness}(a). The fitted experimental profile and the analytical prediction are shown in Fig.~\ref{fig:thickness}(e). For $2\theta=64\pm1^\circ$, the measured thickness profile follows the decrease predicted by the Hasson--Peck model over the measured range.

We also measured the thickness variation using spatial interferometry, as shown schematically in Fig.~\ref{fig:thickness}(f). A diode laser with wavelength $\lambda=652$~nm illuminated the liquid sheet at an incidence angle of $22\pm1^\circ$. The resulting interference pattern was recorded using a CMOS camera (CS165CU1 Zelux, Thorlabs) with a pixel size of $3.45\,\mu\mathrm{m}$. The spatial interference fringes observed on the sheet are shown in Fig.~\ref{fig:thickness}(h).

An intensity line scan was extracted along the sheet-flow direction ($z$), as illustrated in Fig.~\ref{fig:thickness}(g). The scan covered the region from $z=1.5$ to $5.5$~mm. The positions of the interference maxima and minima were used to determine the relative thickness variation along the sheet. The thickness was calculated using
\begin{equation}
d =
-\frac{\lambda}
{2mZ\mu
\sqrt{1-\dfrac{\sin^2\phi}{\mu^2}}},
\label{eq:spatial_thickness}
\end{equation}
where $m$ is the slope obtained from the fringe analysis, $Z$ is the position coordinate, $\mu$ is the refractive index of the liquid, and $\phi$ is the illumination angle
\cite{dombrowski1960some}. The fitted slope was $m=45$~m$^{-1}$. The reconstructed relative thickness profile is shown in Fig.~\ref{fig:thickness}(i).

The spatial-interference analysis was performed over the region $z=1.1$--$5.5$~mm, as shown in Fig.~\ref{fig:thickness}(i). Combining the results from the spectral and spatial measurements, we find that the sheet thickness varies from approximately $4.8$ to $1.1\,\mu\mathrm{m}$ along the measured sheet-flow direction. The spectral method provides the local absolute thickness from the wavelength-dependent reflected spectrum, whereas the spatial method resolves the thickness variation from the spatial fringe pattern. The agreement between the two measurements confirms the nonuniform micrometer-scale thickness of the liquid sheet under the selected operating conditions.

%%%%%%%%%%%%%%%%%%%%%%%%%%%%%%%%%%%%%%%%%%%%%%%%%%%%%%%%%%%%%%%%%%%%%%%%%%%%%%%%%%%%%%%%%%
%%%%%%%%%%%%%%%%%%%%%%%%%%%%%%%%%%%%%%%%%%%%%%%%%%%%%%%%%%%%%%%%%%%%%%%%%%%%%%%%%%%%%%%%%%
%%%%%%%%%%%%%%%%%%%%%%%%%%%%%%%%%%%%%%%%%%%%%%%%%%%%%%%%%%%%%%%%%%%%%%%%%%%%%%%%%%%%%%%%%%
%%%%%%%%%%%%%%%%%%%%%%%%%%%%%%%%%%%%%%%%%%%%%%%%%%%%%%%%%%%%%%%%%%%%%%%%%%%%%%%%%%%%%%%%%%

\section{Liquid sheet as a self-refreshing plasma mirror}
\label{sec:plasma_mirror}

To test the liquid sheet as a target for intense laser--plasma experiments at high repetition rate, we operated it as a plasma mirror \cite{levy2007double,vincenti2014optical}. The sheet, a 60:40 (by volume) ethylene glycol--deionized-water mixture with a typical thickness of $1$--$5~\mu\mathrm{m}$, and with a flow velocity of $\approx13~\mathrm{m\,s^{-1}}$ was operated in a vacuum chamber at $1.5$~mbar maintained by a roughing oil pump. Motorized stages adjusted the edge position relative to the liquid jet, which sets the sheet orientation and dimensions, and a second stage carrying the whole sheet-generation assembly positioned the sheet in the laser focal plane.

\begin{figure}[h!]
    \centering
    \includegraphics[width=1.0\linewidth]{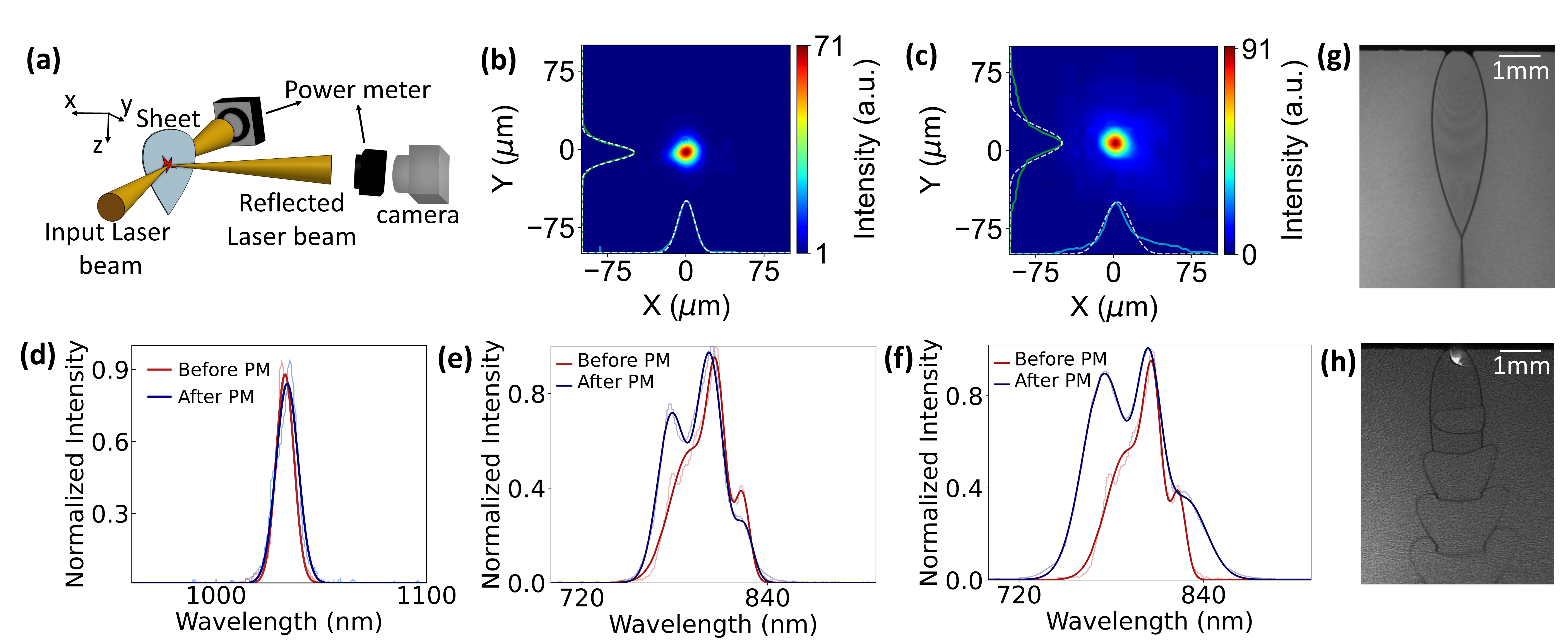}
    \caption{\textbf{Laser interaction with a continuously regenerated liquid-sheet plasma mirror.}
    \textbf{(a)} Experimental configuration for plasma-mirror measurements using a continuously regenerated free-flowing liquid sheet oriented at $45^\circ$ to the incident laser axis.
    \textbf{(b)} Focal-spot profile at the liquid-sheet interaction plane. The measured $1/e^{2}$ beam-waist diameters are $2\omega_x = 33.00~\mu\mathrm{m}$ and $2\omega_y = 32.02~\mu\mathrm{m}$.
    \textbf{(c)} Far-field spatial profile of the reflected Yb:KGW beam, with measured $1/e^{2}$ diameters of $2\omega_x = 51~\mu\mathrm{m}$ and $2\omega_y = 53~\mu\mathrm{m}$.
    \textbf{(d)} Incident and reflected spectra measured with the Yb:KGW laser system (175-fs pulses, $p$ polarization) at a peak intensity of approximately $4\times10^{14}~\mathrm{W\,cm^{-2}}$. The reflected spectrum shows a center-of-mass shift of $+1.0$~nm and an FWHM increase of $2.4$~nm.
    \textbf{(e,f)} Incident and reflected spectra measured with the Ti:sapphire laser system (25-fs pulses, $1$~kHz, $1.8$~mJ) at a peak intensity of approximately $4\times10^{16}~\mathrm{W\,cm^{-2}}$.
    \textbf{(e)} For $s$-polarized incidence, the reflected spectrum shows a center-of-mass shift of $\Delta\lambda_{\mathrm{CoM}}=-3.7$~nm and an FWHM increase from $13.4$ to $20.2$~nm ($\Delta\mathrm{FWHM}=+6.8$~nm).
    \textbf{(f)} For $p$-polarized incidence, the reflected spectrum shows $\Delta\lambda_{\mathrm{CoM}}=-4.7$~nm and an FWHM increase from $13.4$ to $32.8$~nm ($\Delta\mathrm{FWHM}=+19.4$~nm).
    Spectra were recorded with a spectrometer of $0.5$~nm resolution; fiber coupling of the reflected light adds a further, unquantified systematic uncertainty.
    \textbf{(g)} Liquid sheet before laser irradiation.
    \textbf{(h)} Liquid sheet after irradiation at a repetition rate of $10$~kHz, showing regeneration of the interaction surface between successive laser pulses.}
    \label{fig:plasma_mirror}
\end{figure}

The sheet plane was oriented at $45^\circ$ to the laser axis, so that $s$- and $p$-polarized pulses, selected with a half-wave plate, could be compared under identical conditions [Fig.~\ref{fig:plasma_mirror}(a)]. A plano-convex lens ($f=12.5$~cm) focused the beam to a spot with $1/e^2$ diameters of $2\omega_x=33.00~\mu\mathrm{m}$ and $2\omega_y=32.02~\mu\mathrm{m}$ [Fig.~\ref{fig:plasma_mirror}(b)]. The reflected and transmitted beams were collimated with lenses of $f=10$~cm and $3.5$~cm, respectively, and directed out of the chamber through optical windows for diagnostics. Two laser systems were used: a Yb:KGW system (Light Conversion Pharos; $1030$~nm, $175$~fs, $1.5$~mJ, up to $10$~kHz), reaching $I\approx4\times10^{14}~\mathrm{W\,cm^{-2}}$, and a Ti:sapphire system ($\approx800$~nm, $25$~fs, $1.8$~mJ, $1$~kHz), reaching $I\approx4\times10^{16}~\mathrm{W\,cm^{-2}}$. Incident and reflected spectra were recorded with a fiber-coupled spectrometer (Avantes AvaSpec-3648-USB2, resolution $0.5$~nm). 

The far-field profile of the reflected Yb:KGW beam was recorded after the collimating lens, through a neutral-density filter, with a CMOS camera (Blackfly S BFS-U3-51S5M, $3.45~\mu\mathrm{m}$ pixels). It is compact, near-Gaussian and centered on the optical axis, with $1/e^2$ diameters of $51\,\mu\mathrm{m}$ and $53\,\mu\mathrm{m}$ [Fig.~\ref{fig:plasma_mirror}(c)]. This predominantly specular profile~\cite{ziener2003specular,poole2016experiment,streeter2009astra} indicates minimal angular scattering, confirming that the local flatness of the plasma mirror formed on the liquid sheet is preserved within the interaction volume~\cite{Scott_2015}.

The reflected spectrum probes the time-dependent phase imprinted by the plasma mirror, and hence how well the pulse survives reflection. For the Yb:KGW system ($p$ polarization), the spectrum is largely preserved: the center of mass shifts by $+1.0$~nm, which is comparable to the combined resolution and coupling uncertainty and is therefore not considered significant, and the FWHM increases by only $2.4$~nm [Fig.~\ref{fig:plasma_mirror}(d)]. For the Ti:sapphire system, the distortion is considerably stronger and depends on polarization. For $s$ polarization, the spectrum blueshifts by $3.7$~nm and broadens from $13.4$~nm to $20.2$~nm [Fig.~\ref{fig:plasma_mirror}(e)]. For $p$ polarization, the blueshift increases to $4.7$~nm and the FWHM reaches $32.8$~nm [Fig.~\ref{fig:plasma_mirror}(f)]. The distortion thus grows in the order Yb:KGW ($p$) $<$ Ti:sapphire ($s$) $<$ Ti:sapphire ($p$): intensity and pulse duration set the baseline, and polarization amplifies it only at high intensity.

This ordering is consistent with ionization-induced phase modulation. As the leading edge of the pulse field-ionizes the liquid, the rising electron density lowers the refractive index at the reflecting layer, producing a phase $\phi(t)$ and a frequency upshift $\Delta\omega\approx-\partial_t\phi\propto\partial_t n_e$ \cite{landen1992dynamics,le1996spectral}. Because tunnel and multiphoton ionization rates are strongly nonlinear in intensity, the shorter, more intense Ti:sapphire pulse ionizes earlier and faster on its rising edge, giving a larger $\partial_t n_e$ and hence a larger blueshift and broadening. Relativistic effects are negligible ($a_0\ll1$ in both regimes), as is hydrodynamic expansion on the sub-100-fs timescale. The additional enhancement for $p$ polarization is attributed to the field component normal to the surface, which couples energy into the steep density gradient through resonance absorption and Brunel heating \cite{forslund1975theory,brunel1987not,gibbon1992collisionless}. These channels are absent for $s$ polarization and become efficient only at high intensity, which explains why they contribute little at $4\times10^{14}~\mathrm{W\,cm^{-2}}$.

For high-repetition-rate operation, the sheet must replace the laser-affected region between pulses. With a velocity of $13~\mathrm{m\,s^{-1}}$ and a directly affected region of $\approx100~\mu\mathrm{m}$, the geometric replenishment time is $\approx8~\mu\mathrm{s}$ ($\approx120$~kHz); at $10$~kHz the sheet advances $\approx1.2$~mm between pulses. The ruptured film also retracts under surface tension at the Taylor--Culick velocity\cite{savva2009viscous}  $v_{\mathrm{TC}}=\sqrt{2\sigma/\rho h}$, where $\sigma$, $\rho$ and $h$ are the surface tension, density and thickness. For $\sigma\approx50$~mN\,m$^{-1}$, $\rho\approx1.07$~g\,cm$^{-3}$ and $h=2$--$5~\mu\mathrm{m}$, $v_{\mathrm{TC}}\approx4$--$7~\mathrm{m\,s^{-1}}$, so the hole grows by only $\approx0.4$--$0.7$~mm during the $100~\mu\mathrm{s}$ pulse interval, less than the $1.2$~mm advance. The sheet is therefore intact before irradiation and, after $10$~kHz irradiation, shows an expanding plasma plume yet continues to operate as a mirror [Fig.~\ref{fig:plasma_mirror}(g,h)]. The maximum sheet velocity is limited by hydrodynamic instabilities such as fishbone patterns at high flow velocity or Weber \cite{nvkn-hk4g,Cai_Wang_Wu_Chen_2025,10.1063/1.3373513}.

For fixed interaction intensity and pulse duration, the characteristic interaction diameter scales approximately as $d_{\mathrm{int}}\propto\sqrt{E}$. Extrapolating from $30~\mu\mathrm{m}$ at $3$~mJ gives $d_{\mathrm{int}}\approx1.7$--$3.0$~mm for pulse energies of $10$--$30$~J, respectively. At $13~\mathrm{m,s^{-1}}$, the corresponding geometric replenishment times are $\approx0.13$--$0.23$~ms, equivalent to characteristic rates of $\approx4$--$8$~kHz. These estimates indicate that high-energy operation will increasingly depend on the interaction footprint, sheet velocity, and hydrodynamic stability, as well as on the recovery of the irradiated region between pulses.

%%%%%%%%%%%%%%%%%%%%%%%%%%%%%%%%%%%%%%%%%%%%%%%%%%%%%%%%%%%%%%%%%%%%%%%%%%%%%%%%%%%%%%%%%%
%%%%%%%%%%%%%%%%%%%%%%%%%%%%%%%%%%%%%%%%%%%%%%%%%%%%%%%%%%%%%%%%%%%%%%%%%%%%%%%%%%%%%%%%%%
%%%%%%%%%%%%%%%%%%%%%%%%%%%%%%%%%%%%%%%%%%%%%%%%%%%%%%%%%%%%%%%%%%%%%%%%%%%%%%%%%%%%%%%%%%
%%%%%%%%%%%%%%%%%%%%%%%%%%%%%%%%%%%%%%%%%%%%%%%%%%%%%%%%%%%%%%%%%%%%%%%%%%%%%%%%%%%%%%%%%%

\subsection{Reflection, transmission, and absorption of the liquid-sheet plasma mirror}

We investigated the reflection, transmission, and absorption of the liquid-sheet plasma mirror~\cite{doumy2004complete,ziener2003specular,yanovksy1997plasma} using two laser systems with complementary capabilities. For high-repetition-rate operation, we used a Yb:KGW laser system delivering 175-fs pulses at repetition rates up to 10~kHz. Owing to its relatively long pulse duration and available pulse energy of approximately 1.5~mJ, this system provided peak intensities up to approximately $4\times10^{14}~\mathrm{W\,cm^{-2}}$. To investigate the plasma-mirror response at higher intensity, we used a Ti:sapphire, 1-kHz laser system delivering 25-fs pulses with pulse energies of approximately 1--3~mJ. The shorter pulse duration, higher pulse energy, and tighter focusing enabled peak intensities up to approximately $4\times10^{16}~\mathrm{W\,cm^{-2}}$.

\begin{figure}[h!]
    \centering
    \includegraphics[width=\linewidth]{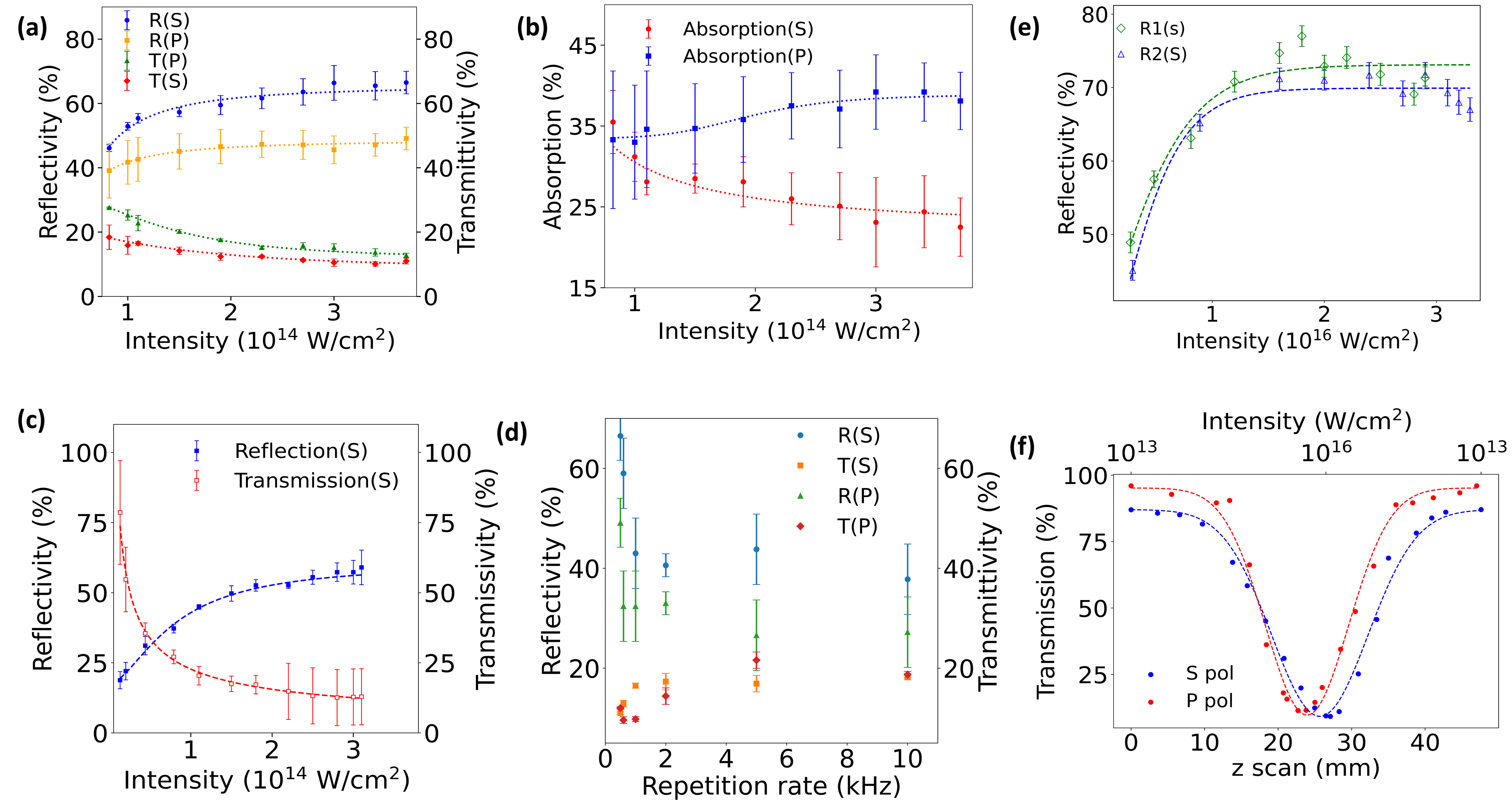}
    \caption{\textbf{Reflection, transmission, and absorption of the liquid-sheet plasma mirror.}
    \textbf{(a)} Reflectivity and transmittivity for $S$- and $P$-polarized light at a repetition rate of 0.5~kHz and a pulse duration of 175~fs.
    \textbf{(b)} Absorptivity inferred from the corresponding measurements in
    (a) using $A=1-R-T$.
    \textbf{(c)} Reflectivity and transmittivity for $S$- and $P$-polarized light at 1~kHz and a pulse duration of 175~fs.
    \textbf{(d)} Reflectivity and transmittivity as functions of repetition rate from 0.5~kHz to 10~kHz for $S$- and $P$-polarized light.
    \textbf{(e)}Intensity-dependent reflectivity of the liquid-sheet plasma mirror measured using a 25-fs Ti:sapphire laser operating at 1~kHz. $R_1$ and $R_2$ denote two independent measurement sets, both acquired with $S$-polarized light. The highest measured reflectivity is approximately 77\%.
    \textbf{(f)} Transmittivity for $S$- and $P$-polarized light as the liquid sheet is translated through the focal region,
    measured using the same 25-fs Ti:sapphire laser at 1~kHz. The upper axis indicates the corresponding incident intensity.}
    \label{fig:reflectivity}
\end{figure}

The experimental arrangement for measuring the reflected and transmitted powers is shown in Fig.~\ref{fig:reflectivity}(a). The reflected and transmitted beams were collimated using lenses with focal lengths of 10~cm and 3.5~cm, respectively, and measured using thermal power-sensor heads (S310C and S425C) connected to a PM100A power-meter console. The incident power on the liquid sheet, $P_{\mathrm{in}}$, was measured separately. The reflectivity, transmittivity, and absorptivity were calculated as
\[
R = \frac{P_{\mathrm{ref}}}{P_{\mathrm{in}}}, \quad
T = \frac{P_{\mathrm{trans}}}{P_{\mathrm{in}}}, \quad
A = 1 - (R + T).
\]

The liquid sheet remained operational at repetition rates up to 10~kHz. During prolonged operation, however, a small amount of laser-induced vapor and droplets gradually deposited on nearby focusing and collimating optics and on the chamber windows. This deposition can reduce optical transmission and modify the beam wavefront, and should therefore be addressed in future implementations using improved shielding or a differentially pumped interaction chamber.

Fig.~\ref{fig:reflectivity}(a) shows the intensity dependence of the liquid-sheet plasma-mirror reflectivity and transmittivity for $S$- and $P$-polarized light at a wavelength of 1030~nm, a pulse duration of 175~fs, and a repetition rate of 0.5~kHz. A reflected signal was first detected at an incident intensity of $8.2\times10^{13}~\mathrm{W\,cm^{-2}}$. As the intensity increased, the reflectivity rose for both polarizations and approached a plateau, whereas the transmittivity decreased. The reflectivity remained higher for $S$ polarization throughout the measured intensity range. At the highest investigated intensity, $3.7\times10^{14}~\mathrm{W\,cm^{-2}}$, the measured reflectivity and transmittivity were $R_S=66.5\%$ and $T_S=11.0\%$, compared with $R_P=49.0\%$ and $T_P=12.8\%$.

Fig.~\ref{fig:reflectivity}(b) shows the absorptivity inferred from the corresponding reflectivity and transmittivity measurements using $A=1-R-T$. The inferred absorptivity exhibits opposite intensity dependences for the two polarizations: it decreases for $S$ polarization but increases modestly for $P$ polarization. At the highest investigated intensity, the inferred values are $A_S=22.5\%$ and $A_P=38.2\%$. Thus, the higher reflectivity obtained with $S$ polarization is accompanied primarily by a lower inferred absorption, rather than by a substantial difference in transmission. The polarization dependence is consistent with polarization-sensitive coupling at an obliquely irradiated plasma surface~\cite{panasenko2010demonstration,inoue2016single,PhysRevAccelBeams.24.121301,10.1063/5.0070963,PhysRevLett.75.252,shaw2016reflectance}. For $P$ polarization, the laser electric field has a component normal to the surface, permitting additional electron-heating channels that are absent for an ideal planar surface under $S$-polarized illumination.

To examine the effect of increasing the repetition rate, we repeated the $S$-polarized reflectivity and transmittivity measurements at 1~kHz, keeping the wavelength and pulse duration unchanged at 1030~nm and 175~fs, respectively. As shown in Fig.~\ref{fig:reflectivity}(c), the reflectivity increases and the transmittivity decreases with incident intensity, following the same overall trends observed at 0.5~kHz [Fig.~\ref{fig:reflectivity}(a)]. Both quantities vary more gradually at higher intensities. The principal difference is a reduction in the measured reflectivity at 1~kHz: the highest value is 59\% at $3.1\times10^{14}~\mathrm{W\,cm^{-2}}$, compared with 66.5\% at a similar incident intensity of $3.7\times10^{14}~\mathrm{W\,cm^{-2}}$ at 0.5~kHz. The corresponding transmittivities remain comparable at 12.9\% and 11.0\%, respectively.

To investigate this repetition-rate dependence systematically, we measured the reflectivity and transmittivity from 0.5 to 10~kHz [Fig.~\ref{fig:reflectivity}(d)]. The laser was operated at 10~kHz, and lower repetition rates were selected using a pulse divider while maintaining a fixed pulse energy. The nominal peak intensity at the liquid sheet was $3.7\times10^{14}~\mathrm{W\,cm^{-2}}$. The measured reflectivity decreases with increasing repetition rate before approaching an approximately constant value. This reduction is attributed primarily to increased laser-generated vapour and progressive deposition of ablated material on optical components inside the vacuum chamber. Deposition on the focusing lens reduces the pulse energy delivered to the liquid sheet, while deposition on the collimating optics in the reflection and transmission arms reduces the optical powers reaching the detectors. The measured repetition-rate dependence therefore reflects the performance of the complete experimental system, including degradation of the illumination and collection paths, rather than the intrinsic plasma-mirror response alone. Despite this reduction, the polarization dependence remained consistent: $S$ polarization produced higher reflectivity and lower transmittivity than $P$ polarization. The liquid sheet continued to form at the interaction location and supported repeated plasma-mirror operation at repetition rates up to 10~kHz.

To extend the characterization to shorter pulses and substantially higher incident intensities, we measured the liquid-sheet plasma-mirror response using a 25-fs Ti:sapphire laser operating at 1~kHz. Figure~\ref{fig:reflectivity}(e) shows two independent reflectivity measurements, denoted $R_1$ and $R_2$, both acquired with $S$-polarized light. In both data sets, the reflectivity rises rapidly at lower intensities and then varies more gradually as the intensity increases. Reflectivities of approximately 65--77\% are obtained over the higher-intensity portion of the scans, with the largest measured value reaching approximately 77\%. At the upper end of the investigated intensity range, both data sets show a modest decrease in reflectivity. The two measurements thus reproduce the same overall intensity dependence, with some variation in the absolute reflectivity.

Whereas pulse-energy variation probes a comparatively narrow intensity range, translating the liquid sheet through the focus changes the illuminated spot area and extends the measurement over approximately three orders of magnitude in intensity, from $10^{13}$ to
$10^{16}~\mathrm{W\,cm^{-2}}$ [Fig.~\ref{fig:reflectivity}(f)]. This $z$-scan connects the low-intensity, predominantly transmitting response of the liquid sheet to its nonlinear plasma-mirror response near focus. Far from focus, the transmission is approximately 85\% for $S$ polarization and 95\% for $P$ polarization, consistent with a largely transparent sheet with polarization-dependent Fresnel losses. As the target approaches focus and the incident intensity increases, the transmission falls sharply to approximately 10\% for both polarizations, before recovering as the target moves beyond focus. 

Together, these measurements demonstrate intensity-dependent switching of the liquid sheet from high transmission to high reflectivity, reaching $77\%$ with 25-fs pulses at 1~kHz. The repetition-rate scans from 0.5 to 10~kHz highlight the concern that vapor and debris accumulate increasingly on the upstream optical components at higher repetition rates. The polarization-resolved measurements show that $s$-polarized pulses exhibit weaker absorption and superior plasma-mirror performance compared with $p$-polarized pulses, making $s$ polarization a suitable choice for driving plasma-mirror operations.

%%%%%%%%%%%%%%%%%%%%%%%%%%%%%%%%%%%%%%%%%%%%%%%%%%%%%%%%%%%%%%%%%%%%%%%%%%%%%%%%%%%%%%%%%%
%%%%%%%%%%%%%%%%%%%%%%%%%%%%%%%%%%%%%%%%%%%%%%%%%%%%%%%%%%%%%%%%%%%%%%%%%%%%%%%%%%%%%%%%%%
%%%%%%%%%%%%%%%%%%%%%%%%%%%%%%%%%%%%%%%%%%%%%%%%%%%%%%%%%%%%%%%%%%%%%%%%%%%%%%%%%%%%%%%%%%
%%%%%%%%%%%%%%%%%%%%%%%%%%%%%%%%%%%%%%%%%%%%%%%%%%%%%%%%%%%%%%%%%%%%%%%%%%%%%%%%%%%%%%%%%%

\subsection{Pointing and intensity stability of the reflected beam}

\begin{figure}[h!]
    \centering
    \includegraphics[width=1\linewidth]{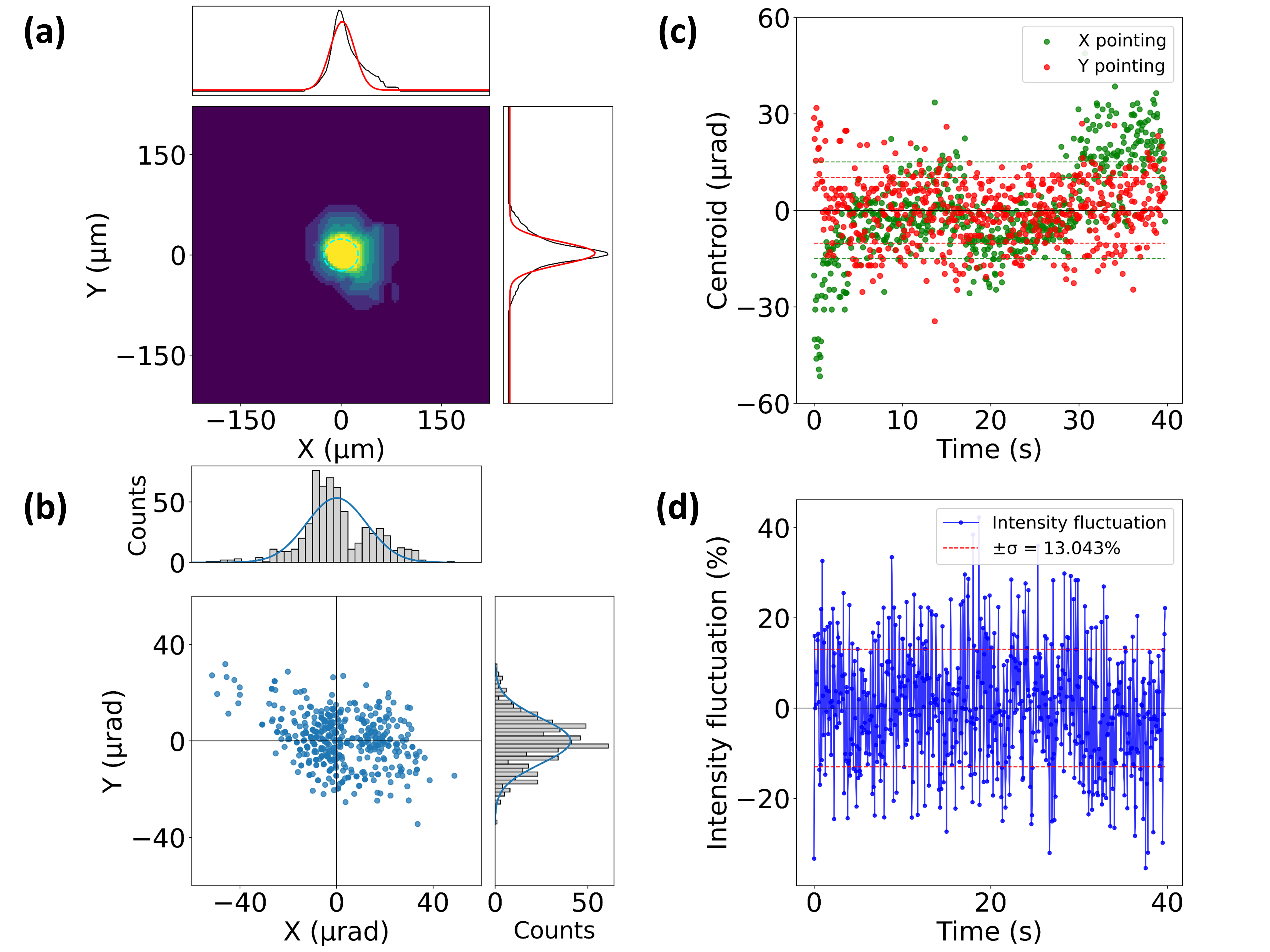}
    \caption{\textbf{Pointing and integrated reflected-intensity stability of the liquid-sheet plasma mirror at 10~kHz.}
    \textbf{(a)} Representative far-field profile of the reflected beam, with diameters $D_x=48\,\mu\mathrm{m}$ and $D_y=43\,\mu\mathrm{m}$.
    \textbf{(b)} Angular centroid positions calculated for $598$ recorded frames. The RMS pointing fluctuations are $15.1\pm0.4\,\mu\mathrm{rad}$ horizontally and $10.2\pm0.3\,\mu\mathrm{rad}$ vertically, corresponding to a quadrature-combined fluctuation of $18.2\pm0.5\,\mu\mathrm{rad}$.
    \textbf{(c)} Temporal evolution of the horizontal and vertical angular centroid positions over approximately $40$~s.
    \textbf{(d)} Temporal evolution of the normalized, frame-integrated reflected-beam intensity during operation at $10$~kHz. The dashed red lines indicate $\pm1\sigma$, where $\sigma=13.0\%$.}
    \label{fig:pointing_stability}
\end{figure}

To assess the pointing stability of the liquid-sheet plasma mirror~\cite{candy2004optimal,fule2024development}, we recorded the reflected far-field beam from the Yb:KGW laser system during operation at $10$~kHz, with an incident pulse energy of $781\,\mu\mathrm{J}$. The liquid sheet was continuously regenerated and ionized in vacuum by successive laser pulses~\cite{thaury2007plasma,doumy2004complete,shaw2016reflectance}. The reflected beam was collimated and focused by a $2$-inch-diameter lens with focal length $f=8$~cm onto a CMOS camera equipped with neutral-density filters. The recorded video contained $598$ frames acquired over approximately $40$~s.

A representative far-field profile is shown in Fig.~\ref{fig:pointing_stability}(a), with beam diameters of $D_x=48\,\mu\mathrm{m}$ and $D_y=43\,\mu\mathrm{m}$. For each frame, we calculated the intensity-weighted beam centroid and converted its focal-plane displacement to angular coordinates using the focal length of the focusing lens. The resulting centroid distribution [Fig.~\ref{fig:pointing_stability}(b)] quantifies the shot-to-shot angular fluctuations of the reflected beam, including possible contributions from mechanical perturbations, liquid-sheet fluctuations, and plasma-induced changes during repeated irradiation.

The RMS centroid fluctuations were $15.1\pm0.4\,\mu\mathrm{rad}$ along $x$ and $10.2\pm0.3\,\mu\mathrm{rad}$ along $y$, giving a quadrature-combined angular fluctuation of $18.2\pm0.5\,\mu\mathrm{rad}$. As shown in Fig.~\ref{fig:pointing_stability}(c), the beam centroid fluctuated around zero along both axes during the $40$-s acquisition, with most measurements lying within approximately $\pm18\,\mu\mathrm{rad}$. Thus, the measured RMS centroid variation provides the angular pointing stability of the reflected beam in the focal plane.

For each frame, the reflected-beam intensity was obtained by integrating the pixel values over the beam profile and normalizing to the mean intensity. As shown in Fig.~\ref{fig:pointing_stability}(d), the normalized intensity fluctuates about unity, with a standard deviation of $13\%$. Individual measurements extend beyond the $\pm\sigma$ limits, indicating frame-to-frame variations in the detected signal during operation at $10$~kHz.

Together with the measured total angular pointing fluctuation of $18.2\pm0.5\,\mu\mathrm{rad}$, these measurements characterize the shot-to-shot reproducibility of the reflected beam during continuous plasma-mirror operation. The limited angular variation supports the use of this target for repeated laser--plasma interactions requiring reproducible beam delivery, including laser-driven secondary-radiation generation~\cite{clark2021high}.

%%%%%%%%%%%%%%%%%%%%%%%%%%%%%%%%%%%%%%%%%%%%%%%%%%%%%%%%%%%%%%%%%%%%%%%%%%%%%%%%%%%%%%%%%%
%%%%%%%%%%%%%%%%%%%%%%%%%%%%%%%%%%%%%%%%%%%%%%%%%%%%%%%%%%%%%%%%%%%%%%%%%%%%%%%%%%%%%%%%%%
%%%%%%%%%%%%%%%%%%%%%%%%%%%%%%%%%%%%%%%%%%%%%%%%%%%%%%%%%%%%%%%%%%%%%%%%%%%%%%%%%%%%%%%%%%
%%%%%%%%%%%%%%%%%%%%%%%%%%%%%%%%%%%%%%%%%%%%%%%%%%%%%%%%%%%%%%%%%%%%%%%%%%%%%%%%%%%%%%%%%%
\section{Operational constraints at high repetition rates}

Scaling laser--liquid-sheet interactions to high repetition rates introduces practical difficulties, mainly because vapor and debris produced at the interaction point contaminate the nearby optical surfaces. The most stable and reproducible plasma-mirror operation was obtained at $0.5$~kHz [Fig.~\ref{fig:reflectivity}(a)]. When the repetition rate was increased to $10$~kHz, both the reflected and transmitted signals degraded [Fig.~\ref{fig:reflectivity}(d)]. This degradation cannot be attributed to a loss of liquid-sheet integrity. In fact, even during 10 kHz operation, the sheet remains continuously renewed, and ablated material is displaced before the subsequent pulse arrives [Fig.~\ref{fig:plasma_mirror}(h)]. The more likely cause is contamination by vapor, fine droplets, and debris generated during the laser--sheet interaction. These deposits affect not only the incident focusing optics but also the collimating optics along both the reflected and transmitted beams. The problem is particularly severe for high-$f$-number optics, where even a small amount of contamination within the pupil can appreciably reduce the optical throughput. Because the deposits accumulate on optical surfaces and vacuum windows, they cannot be removed while the chamber remains under vacuum. Operation must therefore be interrupted to vent the chamber, clean the affected components, and re-establish vacuum. These interruptions limit the duration of continuous operation.

Vapor and liquid handling in vacuum have been addressed in liquid-jet experiments using cryogenic trapping with liquid-nitrogen cold traps~\cite{faubel1988molecular,barnard2022delivery,riley2019design}, heated catchers, differential pumping~\cite{george2019high}, and closed-loop recirculation~\cite{ekimova2015liquid,galinis2017micrometer}. Applying comparable measures here would extend the operating time. In particular, a closed loop returning the collected liquid to the source reservoir under vacuum would prevent accumulation in the catcher, while a catcher connected to a separately or differentially pumped reservoir would improve liquid containment. Protective apertures, shutters, or motorized covers would additionally reduce the exposure of nearby optics and diagnostics to vapor, droplets, and debris.

\section{Conclusion}

This work establishes single-jet impact on a sharp edge as a practical alternative to colliding-jet geometries for generating freely suspended liquid sheets. By systematically examining the impact geometry, flow conditions, and liquid composition, we identify how these parameters govern sheet formation and stability. The interferometric measurements further resolve the spatial nonuniformity of the sheet thickness, while comparison with analytical models clarifies their applicability to this geometry. Beyond sheet generation, the plasma-mirror experiments demonstrate that the continuously renewed surface can sustain repeated laser interactions while maintaining a reproducible reflected-beam direction. The principal challenge for prolonged operation is no longer simply the regeneration of the liquid surface, but the accumulation of laser-generated vapor and droplets on nearby optical components. Overcoming optical contamination through improved shielding and differential pumping will be essential for extending operating times and enabling more reliable operation at higher average powers and repetition rates. The present study establishes a practical foundation for continuously renewed liquid-sheet plasma mirrors and their future application in sustained laser--plasma interactions and secondary-radiation generation.

\section*{Acknowledgments}
This work was supported by intramural funding at TIFR Hyderabad from the Department of Atomic Energy, Government of India, under Project Identification No.~RTI 4007.
%\newpage
%\bibliographystyle{unsrt} % This line ensures references are ordered by citation
%\bibliographystyle{aipnum4-1}
\bibliography{mybib_abbreviated} % Replace myreferences with your .bib file name

@article{gibbon1996short,
  title={Short-pulse laser-plasma interactions},
  author={Gibbon, Paul and F{\"o}rster, Eckhart},
  journal={Plasma Phys. Control. Fusion},
  volume={38},
  number={6},
  pages={769},
  year={1996},
  publisher={IOP Publishing},
 
}

@article{nawani2021review,
  title={A review on multiple liquid jet impingement onto flat plate},
  author={Nawani, Shikshit and Subhash, Maharshi},
  journal={Mater. Today Proc.},
  volume={46},
  pages={11190--11197},
  year={2021},
  publisher={Elsevier}
}

@article{article,
    author = {Hoffman, David and Van Driel, Tim and Kroll, Thomas and Crissman, Christopher and Ryland, Elizabeth and Nelson, Kacie and Cordones, Amy and Koralek, Jake and DePonte, Daniel},
    year = {2022},
    month = {12},
    pages = {1048932},
    title = {Microfluidic liquid sheets as large-area targets for high repetition XFELs},
    volume = {9},
    journal = {Front. Mol. Biosci.},
}

@article{ohba2014effect,
  title={Effect of liquid-sheet thickness on detection sensitivity for laser-induced breakdown spectroscopy of aqueous solution},
  author={Ohba, Hironori and Saeki, Morihisa and Wakaida, Ikuo and Tanabe, Rie and Ito, Yoshiro},
  journal={Opt. Express},
  volume={22},
  number={20},
  pages={24478--24490},
  year={2014},
}

@article{clark2021high,
  title={High-intensity laser-driven secondary radiation sources using the ZEUS 45 TW laser system at the Institute of Plasma Physics and Lasers of the Hellenic Mediterranean University Research Centre},
  author={Clark, EL and Grigoriadis, A and Petrakis, S and Tazes, I and Andrianaki, G and Skoulakis, A and Orphanos, Y and Kaselouris, E and Fitilis, I and Chatzakis, J and others},
  journal={High Power Laser Sci. Eng.},
  volume={9},
  pages={e53},
  year={2021}
}

@article{10.1063/1.5093613,
    author = {Henares, J. L. and Puyuelo-Valdes, P. and Hannachi, F. and Ceccotti, T. and Ehret, M. and Gobet, F. and Lancia, L. and Marqu{\`e}s, J.-R. and Santos, J. J. and Versteegen, M. and Tarisien, M.},
    title = {Development of gas jet targets for laser-plasma experiments at near-critical density},
    journal = {Rev. Sci. Instrum.},
    volume = {90},
    number = {6},
    pages = {063302},
    year = {2019},
    month = {06}
}

@article{ahmed2008characteristics,
  title={Characteristics of liquid sheets formed by splash plate nozzles},
  author={Ahmed, M and Amighi, A and Ashgriz, N and Tran, HN},
  journal={Exp. Fluids},
  volume={44},
  number={1},
  pages={125--136},
  year={2008},
  publisher={Springer}
}

@article{dombrowski1954photographic,
  title={A photographic investigation into the disintegration of liquid sheets},
  author={Dombrowski, Norman and Fraser, Reginald Percy},
  journal={Philos. Trans. R. Soc. Lond. A},
  volume={247},
  number={924},
  pages={101--130},
  year={1954},
  publisher={The Royal Society London}
}

@article{taylor1961formation,
  title={Formation of thin flat sheets of water},
  author={Taylor, Geoffrey},
  journal={Proc. R. Soc. Lond. A},
  volume={259},
  number={1296},
  pages={1--17},
  year={1961},
  publisher={The Royal Society London}
}

@article{choo2002velocity,
  title={The velocity distribution of the liquid sheet formed by two low-speed impinging jets},
  author={Choo, Y-J and Kang, B-S},
  journal={Phys. Fluids},
  volume={14},
  number={2},
  pages={622--627},
  year={2002},
  publisher={American Institute of Physics}
}

@article{sanjay2017formation,
  title={Formation of liquid chain by collision of two laminar jets},
  author={Sanjay, Vatsal and Das, Arup Kumar},
  journal={Phys. Fluids},
  volume={29},
  number={11},
  pages={112101},
  year={2017},
  publisher={AIP Publishing}
}

@article{hasson1964thickness,
  title={Thickness distribution in a sheet formed by impinging jets},
  author={Hasson, David and Peck, Ralph E},
  journal={AIChE J.},
  volume={10},
  number={5},
  pages={752--754},
  year={1964},
  publisher={Wiley Online Library}
}

@article{ibrahim1991impinging,
  title={Impinging jets atomization},
  author={Ibrahim, EA and Przekwas, AJ},
  journal={Phys. Fluids A},
  volume={3},
  number={12},
  pages={2981--2987},
  year={1991},
  publisher={American Institute of Physics}
}

@article{li2006characteristics,
  title={Characteristics of liquid sheets formed by two impinging jets},
  author={Li, Ri and Ashgriz, Nasser},
  journal={Phys. Fluids},
  volume={18},
  number={8},
  pages={087104},
  year={2006},
  publisher={AIP Publishing}
}

@article{yang2014liquid,
  title={Liquid sheet formed by impingement of two viscous jets},
  author={Yang, Li-jun and Zhao, Fei and Fu, Qing-fei and Cui, Kun-da},
  journal={J. Propuls. Power},
  volume={30},
  number={4},
  pages={1016--1026},
  year={2014},
  publisher={American Institute of Aeronautics and Astronautics}
}

@article{hoffman2022liquid,
  title={Liquid heterostructures: Generation of liquid--liquid interfaces in free-flowing liquid sheets},
  author={Hoffman, David J and Bechtel, Hans A and Huyke, Diego A and Santiago, Juan G and DePonte, Daniel P and Koralek, Jake D},
  journal={Langmuir},
  volume={38},
  number={42},
  pages={12822--12832},
  year={2022},
  publisher={ACS Publications}
}

@article{koralek2018generation,
  title={Generation and characterization of ultrathin free-flowing liquid sheets},
  author={Koralek, Jake D and Kim, Jongjin B and Br{\r{u}}{\v{z}}a, Petr and Curry, Chandra B and Chen, Zhijiang and Bechtel, Hans A and Cordones, Amy A and Sperling, Philipp and Toleikis, Sven and Kern, Jan F and others},
  journal={Nat. Commun.},
  volume={9},
  number={1},
  pages={1353},
  year={2018},
  publisher={Nature Publishing Group UK London}
}

@book{stenzel2005physics,
  title={The physics of thin film optical spectra: an introduction},
  author={Stenzel, Olaf},
  year={2005},
  publisher={Springer}
}

@article{kondoh2014liquid,
  title={Liquid-sheet jets for terahertz spectroscopy},
  author={Kondoh, Masato and Tsubouchi, Masaaki},
  journal={Opt. Express},
  volume={22},
  number={12},
  pages={14135--14147},
  year={2014},
  publisher={OSA}
}

@article{watanabe1989new,
  title={A new nozzle producing ultrathin liquid sheets for femtosecond pulse dye lasers},
  author={Watanabe, Akira and Saito, Hiroshi and Ishida, Yuzo and Nakamoto, Masafumi and Yajima, Tatsuo},
  journal={Opt. Commun.},
  volume={71},
  number={5},
  pages={301--304},
  year={1989},
  publisher={Elsevier}
}

@article{galinis2017micrometer,
  title={Micrometer-thickness liquid sheet jets flowing in vacuum},
  author={Galinis, Gediminas and Strucka, Jergus and Barnard, Jonathan CT and Braun, Avi and Smith, Roland A and Marangos, Jon P},
  journal={Rev. Sci. Instrum.},
  volume={88},
  number={8},
  year={2017},
  publisher={AIP Publishing}
}

@article{dombrowski1960some,
  title={Some aspects of liquid flow through fan spray nozzles},
  author={Dombrowski, Norman and Hasson, David and Ward, DE},
  journal={Chem. Eng. Sci.},
  volume={12},
  number={1},
  pages={35--50},
  year={1960},
  publisher={Elsevier}
}

@article{kim2020experimental,
  title={Experimental study on splash phenomena of liquid jet impinging on a vertical wall},
  author={Kim, Hyunsuh and Choi, Hyunhun and Kim, Daegil and Chung, Jaewon and Kim, Hyojun and Lee, Kihyun},
  journal={Exp. Therm. Fluid Sci.},
  volume={116},
  pages={110111},
  year={2020},
  publisher={Elsevier}
}

@article{10.1115/1.2911284,
    author = {Lienhard, J. H., V and Liu, X. and Gabour, L. A.},
    title = {Splattering and Heat Transfer During Impingement of a Turbulent Liquid Jet},
    journal = {J. Heat Transfer},
    volume = {114},
    number = {2},
    pages = {362-372},
    year = {1992},
    month = {05}
}

@article{watson1964radial,
  title={The radial spread of a liquid jet over a horizontal plane},
  author={Watson, EJ},
  journal={J. Fluid Mech.},
  volume={20},
  number={3},
  pages={481--499},
  year={1964},
  publisher={Cambridge University Press}
}

@article{10.1063/5.0047568,
    author = {Mohajer, Behzad and Dey, Anshuman and Li, Ri},
    title = {Stagnation line due to colliding wall jets of two liquid impinging jets on a horizontal surface},
    journal = {Phys. Fluids},
    volume = {33},
    number = {4},
    pages = {047112},
    year = {2021},
    month = {04}
}

@article{taylor1966oblique,
  title={Oblique impact of a jet on a plane surface},
  author={Taylor, Geoffrey},
  journal={Philos. Trans. R. Soc. Lond. A},
  pages={96--100},
  year={1966},
  publisher={JSTOR}
}

@article{jalil2006oblique,
  title={Oblique impingement of circular water jets on a plane boundary},
  author={Jalil, A and Rajaratnam, N},
  journal={J. Hydraul. Res.},
  volume={44},
  number={6},
  pages={807--814},
  year={2006},
  publisher={Taylor \& Francis}
}

@article{kate2007hydraulic,
  title={Hydraulic jumps due to oblique impingement of circular liquid jets on a flat horizontal surface},
  author={Kate, RP and Das, PK and Chakraborty, Suman},
  journal={J. Fluid Mech.},
  volume={573},
  pages={247--263},
  year={2007},
  publisher={Cambridge University Press}
}

@article{vincenti2014optical,
  title={Optical properties of relativistic plasma mirrors},
  author={Vincenti, H and Monchoc{\'e}, S and Kahaly, S and Bonnaud, G and Martin, Ph and Qu{\'e}r{\'e}, F},
  journal={Nat. Commun.},
  volume={5},
  number={1},
  pages={3403},
  year={2014},
  publisher={Nature Publishing Group UK London}
}

@article{poole2016experiment,
  title={Experiment and simulation of novel liquid crystal plasma mirrors for high contrast, intense laser pulses},
  author={Poole, Patrick L and Krygier, Andrew and Cochran, Ginevra E and Foster, PS and Scott, GG and Wilson, LA and Bailey, J and Bourgeois, N and Hernandez-Gomez, Cristina and Neely, David and others},
  journal={Sci. Rep.},
  volume={6},
  number={1},
  pages={32041},
  year={2016},
  publisher={Nature Publishing Group UK London}
}

@book{kruer2019physics,
  title={The physics of laser plasma interactions},
  author={Kruer, William},
  year={2019},
  publisher={crc Press}
}

@article{ziener2003specular,
  title={Specular reflectivity of plasma mirrors as a function of intensity, pulse duration, and angle of incidence},
  author={Ziener, Ch and Foster, PS and Divall, EJ and Hooker, CJ and Hutchinson, MHR and Langley, AJ and Neely, D},
  journal={J. Appl. Phys.},
  volume={93},
  number={1},
  pages={768--770},
  year={2003},
  publisher={American Institute of Physics}
}

@article{10.1063/5.0070963,
    author = {Ponomareva, E. and Ismagilov, A. and Putilin, S. and Tcypkin, A. N.},
    title = {Plasma reflectivity behavior under strong subpicosecond excitation of liquids},
    journal = {APL Photonics},
    volume = {6},
    number = {12},
    pages = {126101},
    year = {2021},
    month = {12}
}

@article{PhysRevAccelBeams.24.121301,
  title = {Emittance preserving thin film plasma mirrors for GeV scale laser plasma accelerators},
  author = {Zingale, A. and Czapla, N. and Nasir, D. M. and Barber, S. K. and Bin, J. H. and Gonsalves, A. J. and Isono, F. and van Tilborg, J. and Steinke, S. and Nakamura, K. and Cochran, G. E. and Purcell, J. and Leemans, W. P. and Geddes, C. G. R. and Schroeder, C. B. and Esarey, E. and Schumacher, D. W.},
  journal = {Phys. Rev. Accel. Beams},
  volume = {24},
  issue = {12},
  pages = {121301},
  numpages = {9},
  year = {2021},
  month = {Dec}
}

@article{underwood2020characterization,
  title={Characterization of flowing liquid films as a regenerating plasma mirror for high repetition-rate laser contrast enhancement},
  author={Underwood, CID and Gan, G and He, Z-H and Murphy, CD and Thomas, AGR and Krushelnick, K and Nees, J},
  journal={Laser Part. Beams},
  volume={38},
  number={2},
  pages={128--134},
  year={2020},
  publisher={Cambridge University Press}
}

@article{doumy2004complete,
  title={Complete characterization of a plasma mirror for the production of high-contrast ultraintense laser pulses},
  author={Doumy, G and Qu{\'e}r{\'e}, F and Gobert, O and Perdrix, M and Martin, Ph and Audebert, P and Gauthier, JC and Geindre, J-P and Wittmann, T},
  journal={Phys. Rev. E},
  volume={69},
  number={2},
  pages={026402},
  year={2004},
  publisher={APS}
}

@article{panasenko2010demonstration,
  title={Demonstration of a plasma mirror based on a laminar flow water film},
  author={Panasenko, Dmitriy and Shu, Anthony J and Gonsalves, Anthony and Nakamura, Kei and Matlis, Nicholas H and Toth, Csaba and Leemans, Wim P},
  journal={J. Appl. Phys.},
  volume={108},
  number={4},
  pages={044913},
  year={2010},
  publisher={AIP Publishing}
}

@article{inoue2016single,
  title={Single plasma mirror providing 104 contrast enhancement and 70\% reflectivity for intense femtosecond lasers},
  author={Inoue, Shunsuke and Maeda, Kazuya and Tokita, Shigeki and Mori, Kazuaki and Teramoto, Kensuke and Hashida, Masaki and Sakabe, Shuji},
  journal={Appl. Opt.},
  volume={55},
  number={21},
  pages={5647--5651},
  year={2016},
  publisher={Optical Society of America}
}

@techreport{yanovksy1997plasma,
  title={Plasma mirrors for short pulse lasers},
  author={Yanovksy, VP and Perry, MD and Brown, CG and Feit, MD and Rubenchik, A},
  year={1997},
  institution={Lawrence Livermore National Lab.(LLNL), Livermore, CA (United States)}
}

@article{shaw2016reflectance,
  title={Reflectance characterization of tape-based plasma mirrors},
  author={Shaw, BH and Steinke, Sven and Van Tilborg, J and Leemans, WP},
  journal={Phys. Plasmas},
  volume={23},
  number={6},
  pages={063118},
  year={2016},
  publisher={AIP Publishing}
}

@article{cao2023vibration,
  title={Vibration and jitter of free-flowing thin liquid sheets as target for high-repetition-rate laser-ion acceleration},
  author={Cao, Zhengxuan and Peng, Ziyang and Shou, Yinren and Zhao, Jiarui and Chen, Shiyou and Gao, Ying and Liu, Jianbo and Wang, Pengjie and Mei, Zhusong and Pan, Zhuo and others},
  journal={Front. Phys.},
  volume={11},
  pages={1172075},
  year={2023},
  publisher={Frontiers Media SA}
}

@article{barnard2022delivery,
  title={Delivery of stable ultra-thin liquid sheets in vacuum for biochemical spectroscopy},
  author={Barnard, Jonathan CT and Lee, Jacob P and Alexander, Oliver and Jarosch, Sebastian and Garratt, Douglas and Picciuto, Rose and Kowalczyk, Katarzyna and Ferchaud, Clement and Gregory, Andrew and Matthews, Mary and others},
  journal={Front. Mol. Biosci.},
  volume={9},
  pages={1044610},
  year={2022},
  publisher={Frontiers Media SA}
}

@article{ha2018device,
  title={Device design and flow scaling for liquid sheet jets},
  author={Ha, Byunghang and DePonte, Daniel P and Santiago, Juan G},
  journal={Phys. Rev. Fluids},
  volume={3},
  number={11},
  pages={114202},
  year={2018},
  publisher={APS}
}

@article{bush2004collision,
  title={On the collision of laminar jets: fluid chains and fishbones},
  author={Bush, John WM and Hasha, Alexander E},
  journal={J. Fluid Mech.},
  volume={511},
  pages={285--310},
  year={2004},
  publisher={Cambridge University Press}
}

@article{levy2007double,
  title={Double plasma mirror for ultrahigh temporal contrast ultraintense laser pulses},
  author={L{\'e}vy, Anna and Ceccotti, Tiberio and d'Oliveira, Pascal and R{\'e}au, Fabrice and Perdrix, Michel and Qu{\'e}r{\'e}, Fabien and Monot, Pascal and Bougeard, Michel and Lagadec, Herv{\'e} and Martin, Philippe and others},
  journal={Opt. Lett.},
  volume={32},
  number={3},
  pages={310--312},
  year={2007},
  publisher={Optical Society of America}
}

@article{thaury2007plasma,
  title={Plasma mirrors for ultrahigh-intensity optics},
  author={Thaury, C{\'e}dric and Quere, Fabien and Geindre, J-P and Levy, Anna and Ceccotti, Tiberio and Monot, P and Bougeard, Michel and R{\'e}au, F and d'Oliveira, P and Audebert, Patrick and others},
  journal={Nat. Phys.},
  volume={3},
  number={6},
  pages={424--429},
  year={2007},
  publisher={Nature Publishing Group UK London}
}

@article{kim2023high,
  title={High-harmonic generation from a flat liquid-sheet plasma mirror},
  author={Kim, Yang Hwan and Kim, Hyeon and Park, Seong Cheol and Kwon, Yongjin and Yeom, Kyunghoon and Cho, Wosik and Kwon, Taeyong and Yun, Hyeok and Sung, Jae Hee and Lee, Seong Ku and others},
  journal={Nat. Commun.},
  volume={14},
  number={1},
  pages={2328},
  year={2023},
  publisher={Nature Publishing Group UK London}
}

@article{streeter2009astra,
  title={Astra gemini compact plasma mirror system},
  author={Streeter, MJV and Foster, PS and Cameron, FH and Bickerton, R and Blake, S and Brummit, P and Costello, B and Divall, E and Hooker, C and Holligan, P and others},
  journal={Cent. Laser Facil. Annu. Rep.},
  pages={225--227},
  year={2009},
}

@article{Scott_2015,
    author = {Scott, G G and Bagnoud, V and Brabetz, C and Clarke, R J and Green, J S and Heathcote, R I and Powell, H W and Zielbauer, B and Arber, T D and McKenna, P and Neely, D},
    year = {2015},
    month = {mar},
    publisher = {IOP Publishing},
    volume = {17},
    number = {3},
    pages = {033027},    
    title = {Optimization of plasma mirror reflectivity and optical quality using double laser pulses},
    journal = {New J. Phys.},
    publisher={Nature Publishing Group UK London}
}

@article{PhysRevLett.75.252,
  title = {Absorption of Ultrashort Laser Pulses by Solid Targets Heated Rapidly to Temperatures 1--1000 eV},
  author = {Price, D. F. and More, R. M. and Walling, R. S. and Guethlein, G. and Shepherd, R. L. and Stewart, R. E. and White, W. E.},
  journal = {Phys. Rev. Lett.},
  volume = {75},
  issue = {2},
  pages = {252--255},
  numpages = {0},
  year = {1995},
  month = {Jul}
}

@article{zhao2016characteristics,
  title={Characteristics of sheet formed by collision of two elliptical jets at short impact distance},
  author={Zhao, Fei and Yang, Li-jun and Mo, Chao-jie and Li, Xue-de},
  journal={J. Fluids Eng.},
  volume={138},
  number={5},
  pages={051201},
  year={2016},
  publisher={American Society of Mechanical Engineers}
}

@article{weng2025experimental,
  title={Experimental study of single and doublet liquid jet impingement on the liquid film of a horizontal heated plate},
  author={Weng, Jung-Yi and Zhan, Ming-Liang and Liu, Yao-Hsien},
  journal={Heat Mass Transfer},
  volume={61},
  number={5},
  pages={44},
  year={2025},
  publisher={Springer}
}

@inproceedings{candy2004optimal,
  title={Optimal Centroid position estimation},
  author={Candy, James V and McClay III, Wilbert A and Awwal, Abdul Ahad Sami and Ferguson, Walter},
  booktitle={Photonic Devices and Algorithms for Computing VI},
  volume={5556},
  pages={249--260},
  year={2004},
  organization={SPIE}
}

@article{fule2024development,
  title={Development of an ultrathin liquid sheet target for laser ion acceleration at high repetition rates in the kilohertz range},
  author={F{\"u}le, M and Kov{\'a}cs, Attila P{\'a}l and Gilinger, T and Karnok, M and Ga{\'a}l, P and Figul, S and Marowsky, G and Osvay, K{\'a}roly},
  journal={High Power Laser Sci. Eng.},
  volume={12},
  pages={e37},
  year={2024},
  publisher={Cambridge University Press}
}

@article{dover2016optical,
  title={Optical shaping of gas targets for laser--plasma ion sources},
  author={Dover, NP and Cook, N and Tresca, O and Ettlinger, O and Maharjan, C and Polyanskiy, MN and Shkolnikov, P and Pogorelsky, I and Najmudin, Z},
  journal={J. Plasma Phys.},
  volume={82},
  number={1},
  pages={415820101},
  year={2016},
  publisher={Cambridge University Press}
}

@article{ekimova2015liquid,
  title={A liquid flatjet system for solution phase soft-x-ray spectroscopy},
  author={Ekimova, Maria and Quevedo, Wilson and Faubel, Manfred and Wernet, Philippe and Nibbering, Erik TJ},
  journal={Struct. Dyn.},
  volume={2},
  number={5},
  pages={054301},
  year={2015},
  publisher={AIP Publishing}
}

@article{menzi2020generation,
  title={Generation and simple characterization of flat, liquid jets},
  author={Menzi, Samuel and Knopp, Gregor and Al Haddad, Andre and Augustin, Sven and Borca, Camelia and Gashi, Dardan and Huthwelker, Thomas and James, Daniel and Jin, Jiaye and Pamfilidis, Georgios and others},
  journal={Rev. Sci. Instrum.},
  volume={91},
  number={10},
  pages={105109},
  year={2020},
  publisher={AIP Publishing}
}

@article{peng2024comprehensive,
  title={A comprehensive diagnostic system of ultra-thin liquid sheet targets},
  author={Peng, Ziyang and Cao, Zhengxuan and Liu, Xuan and Shou, Yinren and Zhao, Jiarui and Chen, Shiyou and Gao, Ying and Wang, Pengjie and Mei, Zhusong and Pan, Zhuo and others},
  journal={High Power Laser Sci. Eng.},
  volume={12},
  pages={e26},
  year={2024},
  publisher={Cambridge University Press}
}

@article{gallo2024development,
  title={Development of a flat jet delivery system for soft X-ray spectroscopy at MAX IV},
  author={Gallo, Tamires and Adriano, Luigi and Heymann, Michael and Wrona, Agnieszka and Walsh, Noelle and {\"O}hrwall, Gunnar and Callefo, Flavia and Skruszewicz, Slawomir and Namboodiri, Mahesh and Marinho, Ricardo and others},
  journal={Synchrotron Radiat.},
  volume={31},
  number={5},
  pages={1285--1292},
  year={2024},
  publisher={International Union of Crystallography}
}

@article{crissman2022sub,
  title={Sub-micron thick liquid sheets produced by isotropically etched glass nozzles},
  author={Crissman, Christopher J and Mo, Mianzhen and Chen, Zhijiang and Yang, Jie and Huyke, Diego A and Glenzer, Siegfried H and Ledbetter, Kathryn and F. Nunes, J Pedro and Ng, May Ling and Wang, Hengzi and others},
  journal={Lab Chip},
  volume={22},
  number={7},
  pages={1365--1373},
  year={2022},
  publisher={The Royal Society of Chemistry}
}

@article{condamine2021high,
  title={High-repetition rate solid target delivery system for PW-class laser--matter interaction at ELI Beamlines},
  author={Condamine, FP and Jourdain, N and Hernandez, J-C and Taylor, M and Bohlin, H and Fajstavr, A and Jeong, TM and Kumar, D and La{\v{s}}tovi{\v{c}}ka, T and Renner, O and others},
  journal={Rev. Sci. Instrum.},
  volume={92},
  number={6},
  pages={063504},
  year={2021},
  publisher={AIP Publishing}
}

@article{prencipe2017targets,
  title={Targets for high repetition rate laser facilities: needs, challenges and perspectives},
  author={Prencipe, I and Fuchs, J and Pascarelli, S and Schumacher, DW and Stephens, RB and Alexander, NB and Briggs, R and B{\"u}scher, M and Cernaianu, MO and Choukourov, A and others},
  journal={High Power Laser Sci. Eng.},
  volume={5},
  pages={e17},
  year={2017},
  publisher={Cambridge University Press}
}

@article{xu2023versatile,
  title={Versatile tape-drive target for high-repetition-rate laser-driven proton acceleration},
  author={Xu, Nuo and Streeter, Matthew JV and Ettlinger, OC and Ahmed, Hamad and Astbury, Sam and Borghesi, M and Bourgeois, N and Curry, CB and Dann, SJD and Dover, NP and others},
  journal={High Power Laser Sci. Eng.},
  volume={11},
  pages={e23},
  year={2023},
  publisher={Cambridge University Press}
}

@article{knight2024detailed,
  title={Detailed characterization of kHz-rate laser-driven fusion at a thin liquid sheet with a neutron detection suite},
  author={Knight, Benjamin M and Gautam, Connor M and Stoner, Colton R and Egner, Bryan V and Smith, Joseph R and Orban, Chris M and Manfredi, Juan J and Frische, Kyle D and Dexter, Michael L and Chowdhury, Enam A and others},
  journal={High Power Laser Sci. Eng.},
  volume={12},
  pages={e2},
  year={2024},
  publisher={Cambridge University Press}
}

@article{hah2016high,
  title={High repetition-rate neutron generation by several-mJ, 35 fs pulses interacting with free-flowing D2O},
  author={Hah, Jungmoo and Petrov, GM and Nees, JA and He, Z-H and Hammig, MD and Krushelnick, Karl and Thomas, AGR},
  journal={Appl. Phys. Lett.},
  volume={109},
  number={14},
  pages={144102},
  year={2016},
  publisher={AIP Publishing}
}

@article{kar2008plasma,
  title={Plasma jets driven by ultraintense-laser interaction with thin foils},
  author={Kar, S and Borghesi, M and Bulanov, SV and Key, MH and Liseykina, TV and Macchi, A and Mackinnon, AJ and Patel, PK and Romagnani, L and Schiavi, Angelo and others},
  journal={Phys. Rev. Lett.},
  volume={100},
  number={22},
  pages={225004},
  year={2008},
  publisher={APS}
}

@article{morrison2018mev,
  title={MeV proton acceleration at kHz repetition rate from ultra-intense laser liquid interaction},
  author={Morrison, John T and Feister, Scott and Frische, Kyle D and Austin, Drake R and Ngirmang, Gregory K and Murphy, Neil R and Orban, Chris and Chowdhury, Enam A and Roquemore, WM},
  journal={New J. Phys.},
  volume={20},
  number={2},
  pages={022001},
  year={2018},
  publisher={IOP Publishing}
}

@article{luu2018extreme,
  title={Extreme--ultraviolet high--harmonic generation in liquids},
  author={Luu, Tran Trung and Yin, Zhong and Jain, Arohi and Gaumnitz, Thomas and Pertot, Yoann and Ma, Jun and W{\"o}rner, Hans Jakob},
  journal={Nat. Commun.},
  volume={9},
  number={1},
  pages={3723},
  year={2018},
  publisher={Nature Publishing Group UK London}
}

@article{souckova2008measurement,
  title={Measurement and correlation of the surface tension- temperature relation for methanol},
  author={Souckova, Monika and Klomfar, Jaroslav and P{\'a}tek, Jaroslav},
  journal={J. Chem. Eng. Data},
  volume={53},
  number={9},
  pages={2233--2236},
  year={2008},
  publisher={ACS Publications}
}

@article{xiang2006new,
  title={A new reference correlation for the viscosity of methanol},
  author={Xiang, Hong Wei and Laesecke, Arno and Huber, Marcia L},
  journal={J. Phys. Chem. Ref. Data},
  volume={35},
  number={4},
  pages={1597--1620},
  year={2006},
  publisher={AIP Publishing}
}

@article{vargaftik1983international,
  title={International tables of the surface tension of water},
  author={Vargaftik, NB and Volkov, Bc Nikolay and Voljak, LD},
  journal={J. Phys. Chem. Ref. Data},
  volume={12},
  number={3},
  pages={817--820},
  year={1983},
  publisher={American Institute of Physics for the National Institute of Standards and~\ldots{}}
}

@article{quijada2013viscosity,
  title={Viscosity and density data for the ternary system water (1)--ethanol (2)--ethylene glycol (3) between 298.15 K and 328.15 K},
  author={Quijada-Maldonado, E and Meindersma, G Wytze and De Haan, AB},
  journal={J. Chem. Thermodyn.},
  volume={57},
  pages={500--505},
  year={2013},
  publisher={Elsevier}
}

@article{schumacher2017liquid,
  title={Liquid crystal targets and plasma mirrors for laser based ion acceleration},
  author={Schumacher, DW and Poole, PL and Willis, C and Cochran, GE and Daskalova, R and Purcell, J and Heery, R},
  journal={J. Instrum.},
  volume={12},
  number={04},
  pages={C04023--C04023},
  year={2017}
}

@article{landen1992dynamics,
  title={Dynamics of picosecond-laser-pulse plasmas determined from the spectral shifts of reflected probe pulses},
  author={Landen, OL and Alley, WE},
  journal={Phys. Rev. A},
  volume={46},
  number={8},
  pages={5089},
  year={1992},
  publisher={APS},
}

@article{le1996spectral,
  title={Spectral, temporal, and spatial characteristics of plasma-induced spectral blue shifting and its application to femtosecond pulse measurement},
  author={Le Blanc, SP and Sauerbrey, R},
  journal={J. Opt. Soc. Am. B},
  volume={13},
  number={1},
  pages={72--88},
  year={1996},
  publisher={Optical Society of America}
}

@article{forslund1975theory,
  title={Theory and simulation of resonant absorption in a hot plasma},
  author={Forslund, DW and Kindel, JM and Lee, Kenneth and Lindman, EL and Morse, RL},
  journal={Phys. Rev. A},
  volume={11},
  number={2},
  pages={679},
  year={1975},
  publisher={APS}
}

@article{brunel1987not,
  title={Not-so-resonant, resonant absorption},
  author={Brunel, F},
  journal={Phys. Rev. Lett.},
  volume={59},
  number={1},
  pages={52},
  year={1987},
  publisher={APS}
}

@article{gibbon1992collisionless,
  title={Collisionless absorption in sharp-edged plasmas},
  author={Gibbon, Paul and Bell, AR},
  journal={Phys. Rev. Lett.},
  volume={68},
  number={10},
  pages={1535},
  year={1992},
  publisher={APS}
}

@article{10.1063/5.0328950,
    author = {Dexter, Michael L. and Hageman, Stephen J. and Ngirmang, Gregory K. and Frische, Kyle D. and Snyder, Joseph and Morrison, John T. and Chowdhury, Enam A. and Patnaik, Anil K.},
    title = {Generation of kHz-rate complex-structured liquid targets for relativistic laser--plasma interactions},
    journal = {Appl. Phys. Lett.},
    volume = {128},
    number = {25},
    pages = {251108},
    year = {2026},
    month = {06}
    }

@article{ggsf-hkm1,
  title = {Characterization and automated optimization of laser-driven proton beams from converging liquid sheet jet targets},
  author = {Glenn, G. D. and Treffert, F. and Ahmed, H. and Astbury, S. and Borghesi, M. and Bourgeois, N. and Curry, C. B. and Dann, S. J. D. and DiIorio, S. and Dover, N. P. and Dzelzainis, T. and Ettlinger, O. and Gauthier, M. and Giuffrida, L. and Gray, R. J. and Green, J. S. and Hicks, G. S. and Hyland, C. and Istokskaia, V. and King, M. and Loughran, B. and Margarone, D. and McCusker, O. and McKenna, P. and Najmudin, Z. and Parisua{\~n}a, C. and Parsons, P. and Spindloe, C. and Streeter, M. J. V. and Symes, D. R. and Thomas, A. G. R. and Xu, N. and Glenzer, S. H. and Palmer, C. A. J.},
  journal = {Phys. Rev. Research},
  volume = {8},
  issue = {1},
  pages = {013101},
  numpages = {17},
  year = {2026},
  month = {Jan}
}

@article{10.1063/5.0316700,
    author = {Tong, Tiantian and Liu, Ye and Hu, Jie and Tang, Guoqiang and Tian, Shan Xi},
    title = {A microfluidic liquid sheet apparatus for extreme ultraviolet high-order harmonic generation},
    journal = {Rev. Sci. Instrum.},
    volume = {97},
    number = {7},
    pages = {073101},
    year = {2026},
    month = {07}
}

@article{10.1063/5.0279613,
    author = {Ivanov, K. A. and Shulyapov, S. A. and Tsymbalov, I. N. and Tsygvintsev, I. P. and Vichev, I. Yu. and Krivokorytov, M. S. and Volkov, R. V. and Savel'ev, A. B.},
    title = {Optically shaped mass-limited liquid sheet target for high repetition rate laser-driven acceleration},
    journal = {Phys. Plasmas},
    volume = {32},
    number = {8},
    pages = {080702},
    year = {2025},
    month = {08},
}

@article{george2019high,
  title={High-repetition-rate (kHz) targets and optics from liquid microjets for high-intensity laser--plasma interactions},
  author={George, KM and Morrison, JT and Feister, S and Ngirmang, GK and Smith, JR and Klim, AJ and Snyder, J and Austin, D and Erbsen, W and Frische, KD and others},
  journal={High Power Laser Sci. Eng.},
  volume={7},
  pages={e50},
  year={2019},
  publisher={Cambridge University Press}
}

@article{treffert2022high,
  title={High-repetition-rate, multi-MeV deuteron acceleration from converging heavy water microjets at laser intensities of 1021 W/cm2},
  author={Treffert, Franziska and Curry, Chandra B and Chou, H-GJ and Crissman, Christopher J and DePonte, Daniel P and Fiuza, Frederico and Glenn, Griffin D and Hollinger, Reed C and Nedbailo, Ryan and Park, Jaebum and others},
  journal={Appl. Phys. Lett.},
  volume={121},
  number={7},
  pages={074104},
  year={2022},
  publisher={AIP Publishing}
}

@article{streeter2025stable,
  title={Stable laser-acceleration of high-flux proton beams with plasma collimation},
  author={Streeter, MJV and Glenn, GD and DiIorio, S and Treffert, F and Loughran, B and Ahmed, H and Astbury, S and Borghesi, M and Bourgeois, N and Curry, CB and others},
  journal={Nat. Commun.},
  volume={16},
  number={1},
  pages={1004},
  year={2025},
  publisher={Nature Publishing Group UK London}
}

@article{ehret2024stability,
  title={Stability and debris-mitigation of a solid tape target delivery system for intense laser-matter interactions towards high-repetition-rate},
  author={Ehret, Michael and de Luis, Diego and Api{\~n}aniz, Jon Imanol and Henares, Jose Luis and Lera, Roberto and P{\'e}rez-Hern{\'a}ndez, Jos{\'e} Antonio and Puyuelo-Valdes, Pilar and Volpe, Luca and Gatti, Giancarlo},
  journal={Plasma Phys. Control. Fusion},
  volume={66},
  number={4},
  pages={045003},
  year={2024},
  publisher={IOP Publishing}
}

@article{bremond2006atomization,
  title={Atomization by jet impact},
  author={Bremond, Nicolas and Villermaux, Emmanuel},
  journal={J. Fluid Mech.},
  volume={549},
  pages={273--306},
  year={2006},
  publisher={Cambridge University Press}
}

@article{Borghesi01042006,
    author = {M. Borghesi and J. Fuchs and S. V. Bulanov and A. J. MacKinnon and P. K. Patel and M. Roth},
    title = {Fast Ion Generation by High-Intensity Laser Irradiation of Solid Targets and Applications},
    journal = {Fusion Sci. Technol.},
    volume = {49},
    number = {3},
    pages = {412--439},
    year = {2006},
    publisher = {Taylor \& Francis},
}

@article{borot2014high,
  title={High repetition rate plasma mirror device for attosecond science},
  author={Borot, A and Douillet, D and Iaquaniello, G and Lefrou, T and Audebert, P and Geindre, J-P and Lopez-Martens, R},
  journal={Rev. Sci. Instrum.},
  volume={85},
  number={1},
  pages={013104},
  year={2014},
  publisher={AIP Publishing}
}

@article{yan1997heat,
  title={Heat transfer from an obliquely impinging circular, air jet to a flat plate},
  author={Yan, Xiaojun and Saniei, Nader},
  journal={Int. J. Heat Fluid Flow},
  volume={18},
  number={6},
  pages={591--599},
  year={1997},
  publisher={Elsevier}
}

@article{Kaewchoothong31052026,
author = {Natthaporn Kaewchoothong},
title = {Influence of Obliquely Spiral Jet Impingement on Flow and Heat Transfer Characteristics of Flat Surface},
journal = {Heat Transfer Eng.},
volume = {47},
number = {10},
pages = {943--960},
year = {2026},
publisher = {Taylor \& Francis},
}

@article{mishra2020experimental,
  title={Experimental study of flow characteristics of an oblique impinging jet},
  author={Mishra, Abhishek and Yadav, Harekrishna and Djenidi, Lyazid and Agrawal, Amit},
  journal={Exp. Fluids},
  volume={61},
  number={3},
  pages={90},
  year={2020},
  publisher={Springer}
}

@article{faubel1988molecular,
  title={A molecular beam study of the evaporation of water from a liquid jet},
  author={Faubel, M and Schlemmer, S and Toennies, JP},
  journal={Z. Phys. D Atoms Mol. Clusters},
  volume={10},
  number={2},
  pages={269--277},
  year={1988},
  publisher={Springer}
}

@article{riley2019design,
  title={Design and characterization of a recirculating liquid-microjet photoelectron spectrometer for multiphoton ultraviolet photoelectron spectroscopy},
  author={Riley, Jamie W and Wang, Bingxing and Parkes, Michael A and Fielding, Helen H},
  journal={Rev. Sci. Instrum.},
  volume={90},
  number={8},
  pages={083104},
  year={2019},
  publisher={AIP Publishing}
}

@article{nvkn-hk4g,
  title = {Two instabilities in one liquid sheet},
  author = {Dighe, Sandip and Gadgil, Hrishikesh and Truscott, Tadd},
  journal = {Phys. Rev. Fluids},
  volume = {10},
  issue = {11},
  pages = {110501},
  numpages = {6},
  year = {2025},
  month = {Nov},
  publisher = {American Physical Society},
}

@article{Cai_Wang_Wu_Chen_2025,
  title={Origin and bifurcations in fishbone instabilities of impinging jets},
  author={Cai, Guohui and Wang, Bo and Wu, Erjun and Chen, Xiaodong},
  journal={J. Fluid Mech.},
  volume={1016},
  pages={A60},
  year={2025},
}

@article{10.1063/1.3373513,
    author = {Jung, Sungjune and Hoath, Stephen D. and Martin, Graham D. and Hutchings, Ian M.},
    title = {Atomization patterns produced by the oblique collision of two Newtonian liquid jets},
    journal = {Phys. Fluids},
    volume = {22},
    number = {4},
    pages = {042101},
    year = {2010},
    month = {04}
    }

@article{10.1039/d1lc00757b,
    author = {Crissman, Christopher J. and Mo, Mianzhen and Chen, Zhijiang and Yang, Jie and Huyke, Diego A. and Glenzer, Siegfried H. and Ledbetter, Kathryn and F. Nunes, J. Pedro and Ng, May Ling and Wang, Hengzi and Shen, Xiaozhe and Wang, Xijie and DePonte, Daniel P.},
    title = {Sub-micron thick liquid sheets produced by isotropically etched glass nozzles},
    journal = {Lab Chip},
    volume = {22},
    number = {7},
    pages = {1365-1373},
    year = {2022},
    month = {04}
    }

@article{puyuelo2022implementation,
  title={Implementation of a thin, flat water target capable of high-repetition-rate MeV-range proton acceleration in a high-power laser at the CLPU},
  author={Puyuelo-Valdes, Pilar and De Luis, D and Hernandez, J and Api{\~n}aniz, JI and Curcio, A and Henares, JL and Huault, M and P{\'e}rez-Hern{\'a}ndez, JA and Roso, L and Gatti, G and others},
  journal={Plasma Phys. Control. Fusion},
  volume={64},
  number={5},
  pages={054003},
  year={2022},
  publisher={IOP Publishing}
}

@article{ngirmang2020evidence,
  title={Evidence of radial Weibel instability in relativistic intensity laser-plasma interactions inside a sub-micron thick liquid target},
  author={Ngirmang, Gregory K and Morrison, John T and George, Kevin M and Smith, Joseph R and Frische, Kyle D and Orban, Chris and Chowdhury, Enam A and Roquemore, W Mel},
  journal={Sci. Rep.},
  volume={10},
  number={1},
  pages={9872},
  year={2020},
  publisher={Nature Publishing Group UK London}
}

@article{PhysRevResearch.7.023190,
  title = {Stable high-energy proton acceleration with water-leaf targets driven by intense laser pulses},
  author = {He, L. R. and Bachhammer, M. and Balling, F. and Biswas, S. and Doyle, L. and Gerlach, S. and Hofrichter, I. and Kharbedia, M. and Liese, J. and De Marco, M. and Pohle, T. and Pra{\ss{}}elsperger, A. and Schmidt, A.-K. and Schweiger, F. and Kling, M. F. and Karsch, S. and Schreiber, J.},
  journal = {Phys. Rev. Research},
  volume = {7},
  issue = {2},
  pages = {023190},
  numpages = {9},
  year = {2025},
  month = {May},
  publisher = {American Physical Society},
}

@article{choo2001parametric,
  title={Parametric study on impinging-jet liquid sheet thickness distribution using an interferometric method},
  author={Choo, YJ and Kang, BS},
  journal={Exp. Fluids},
  volume={31},
  number={1},
  pages={56--62},
  year={2001},
  publisher={Springer}
}

@article{savva2009viscous,
  title={Viscous sheet retraction},
  author={Savva, Nikos and Bush, John WM},
  journal={J. Fluid Mech.},
  volume={626},
  pages={211--240},
  year={2009},
  publisher={Cambridge University Press}
}

\end{document}